\documentclass[11pt]{article}

\usepackage[utf8]{inputenc}
\usepackage[T1]{fontenc}
\usepackage[a4paper, margin=1in]{geometry}
\usepackage{lmodern} 
\usepackage{microtype}
\usepackage{csquotes}
\usepackage{amsmath, amssymb}
\usepackage{hyperref}
\usepackage{float}
\usepackage{graphicx}
\usepackage{longtable}
\usepackage{footnotehyper}
\makesavenoteenv{longtable}
\usepackage{caption}
\usepackage{subcaption}
\usepackage[
    backend=biber,
    style=authoryear
]{biblatex}

\title{\vspace{-2em}\textbf{A Fast Quasi-Linear Heuristic for the Close-Enough Traveling Salesman Problem}\vspace{-1em}}
\date{} 
\author{
Khoi Duong \\
University of Minnesota \\
\texttt{duong245@umn.edu}
}

\begin{document}
\maketitle
\vspace{-2em}

\begin{abstract}
We present a randomized, expected-$O(n\log n)$ construction heuristic for the Close-Enough Traveling Salesman Problem (CETSP). The method combines hierarchical circle clustering with incremental reconstruction and localized geometric improvement. On the canonical Mennell benchmark, a single construction run is approximately 220 times faster than the state-of-the-art MA-CETSP solver \parencite{LeiHao2024} at the median after hardware adjustment, while the best of 1,000 construction runs has a median gap of \(0.35\%\) from the values reported by \textcite{LeiHao2024}. Since our algorithm is fast, we also use it as an initializer for the state-of-the-art MA-CETSP solver. On the 62 canonical Mennell benchmark instances, the best of 20 independent hybrid runs improves the best values reported by \textcite{LeiHao2024} on 17 instances and matches 31 more at the reported precision. These results show that the proposed construction method is useful both as a scalable standalone heuristic and as an effective initializer for higher-cost CETSP optimization.
\end{abstract}

\section{Introduction}

The \emph{Close-Enough Traveling Salesman Problem} (CETSP) is a continuous generalization of the classical Traveling Salesman Problem (TSP), where each target is associated with a circular neighborhood that the salesman must intersect rather than a precise point. 

Concretely, let the instance be given by centers \(c_i\in\mathbb{R}^2\) and radii \(r_i\ge0\) for i=1,\dots,n, and denote the closed target disks
\[
D_i := \{x\in\mathbb{R}^2:\|x-c_i\|\le r_i\},\qquad i=1,\dots,n.
\]
A feasible (discrete) tour can be described by a permutation \(\sigma\in S_n\) and visit points \(p_1,\dots,p_n\) with \(p_i\in D_{\sigma(i)}\). The Close-Enough TSP then seeks
\[
\min_{\sigma\in S_n}\;\min_{p_i\in D_{\sigma(i)}\ (i=1,\dots,n)}
\; \sum_{i=1}^n \|p_i-p_{i+1}\|,\qquad p_{n+1}:=p_1,
\]
i.e. the shortest closed polygonal tour that visits one point inside each disk. (Note that some visit points may coincide, so a single geometric point can satisfy multiple disks.)

This problem generalizes the Euclidean TSP (take all \(r_i=0\)) and is NP-hard in general; the algorithmic task is therefore to produce high-quality approximations or heuristics for the above optimization problem.

The CETSP thus combines the discrete optimization of the visiting sequence with the continuous optimization of visiting positions, and it naturally models path-planning tasks such as radio meter reading, drone-based inspection, and robotic welding \parencite{LeiHao2024,DiPlacido2022}.

Recent high-performing CETSP algorithms rely on population-based (genetic) metaheuristics that achieve strong solution quality at the expense of high computational cost. In this work, we present a new heuristic for the CETSP that is inspired by the quasi-linear \emph{pair-center algorithm} for the Euclidean TSP proposed by \textcite{Formella2024}. While originally designed for point-based TSP instances, the pair-center concept extends naturally to the CETSP once disk geometry and intersection logic are handled properly. The resulting constructor runs in expected $O(n\log n)$ time and can be used either as a fast standalone solver or as a source of high-quality, structurally varied initial tours for a slower memetic solver. The latter hybrid configuration improves 17 of the best values reported by \textcite{LeiHao2024}.

For reproducibility, the code and data for our method are available in \href{https://anonymous.4open.science/r/cetsp-56F6}{this repository}.

\section{Related Work}
The CETSP problem was introduced by \textcite{Gulczynski2006} and has been a subject of research interest in the last two decades. \textcite{Gulczynski2006} introduced a three-step process: first, a set of Steiner zones is constructed that includes all discs; second, these Steiner zones are discretized into points and a TSP tour is constructed over them; third, the tour is locally optimized by improving the positions of these points. \textcite{Mennell2009,Mennell2011} followed this and introduced a Second-Order Cone Programming (SOCP) formulation for the local optimization step, which they solved using commercial solvers. \textcite{Mennell2009} also introduced a dataset that has been widely used for benchmarking CETSP algorithms.

The first exact solver for the CETSP was proposed by \textcite{Behdani2014}, who used a discretization scheme to formulate the problem as a Mixed-Integer Programming (MIP) problem. \textcite{Carrabs2017a} presented a discretization scheme to convert the problem to a Generalized TSP (GTSP), which is then solved using MIP; this was improved on in \textcite{Carrabs2017b}. \textcite{Coutinho2016} introduced an approach based on branch-and-bound and SOCP, which can reach some optimal solutions in a finite number of steps. \textcite{Zhang2023} presents an improved branch-and-bound algorithm that incorporates a number of optimizations. 

Several heuristic approaches have also been proposed. \textcite{Wang2019} presented the SZVNS heuristic, based on the Variable Neighborhood Search. \textcite{Carrabs2020} proposed a heuristic based on the discretization of \textcite{Carrabs2017b} and a carousel greedy algorithm.

The most efficient approximate algorithms for the CETSP currently are evolutionary algorithms. \textcite{DiPlacido2022} proposed a genetic algorithm which incorporates various strategies to optimize the tour and uses the SOCP for tour point optimization. \textcite{Cariou2023} presented another genetic algorithm which incorporated some new geometric heuristics. Most recently, \textcite{LeiHao2024} introduced a new genetic algorithm that includes a new crossover operator which more efficiently combines existing tours. This approach, in particular, statistically outperforms all prior reference algorithms. When evaluated on the \textcite{Mennell2009} dataset, Lei and Hao's algorithm successfully found all 23 known optimal values and set 30 new best upper bounds on the remaining instances, proving highly competitive against all existing state-of-the-art methods. However, these algorithms are inherently computationally expensive (e.g. \textcite{LeiHao2024} reports runtimes of up to 40 minutes for problems of size $n = 1000$), which limits their scalability to yet larger instances.

\section{Algorithm}

Our approach follows the overall structure of the quasi-linear \emph{pair-center algorithm} introduced by \textcite{Formella2024} for the Euclidean TSP, but extends it to handle circular neighborhoods as in the Close-Enough Traveling Salesman Problem. Conceptually, the method retains the same two-phase organization: a \emph{clustering phase} that builds a hierarchical representation of the instance, and a \emph{construction phase} that incrementally expands a feasible tour from this hierarchy.

In the clustering phase, the algorithm recursively merges the closest pair of geometric objects into a new composite ``proxy'' object until a single hierarchy remains. In the CETSP setting, these objects are circles rather than points as in \textcite{Formella2024}. The result is a binary clustering tree whose internal nodes represent proxy circles.

The construction phase then traverses this tree to build a closed tour. As in the original pair-center approach, the tour is dynamically maintained so that it remains feasible at every step. However, the continuous nature of the CETSP introduces two additional challenges: first, multiple circles may correspond to a single effective tour point when their feasible regions overlap; second, each tour point admits continuous local optimization within its circle. To address these, the algorithm performs lightweight dynamic updates---reinserting or locally re-optimizing tour points---while preserving near-linear expected runtime.

Overall, the method retains the speed and structural simplicity of the pair-center algorithm while incorporating the geometric flexibility required for close-enough constraints. Detailed definitions, data structures, and optimization steps are presented in the following sections.

\subsection{Data Structures}

We make extensive use of geometric search structures to maintain and query spatial relationships efficiently. In particular, we employ an R*-tree to store bounding boxes of the current set of circles, enabling logarithmic-time expected queries for nearest neighbors, insertions, and deletions. The resulting operations have expected $O(n \log n)$ runtime; see \textcite{Formella2024} for a detailed review and analysis.

\subsection{Clustering Phase}

The clustering phase closely follows the structure of the pair-center algorithm described by \textcite{Formella2024}, with appropriate modifications to handle circular regions instead of point targets. As in the original approach, the goal is to construct a hierarchical binary tree over all targets by iteratively merging the closest pair of geometric objects, replacing them with a single representative (a \emph{proxy circle}). The leaves of the resulting tree correspond to the original input disks, and each internal node stores its proxy circle and the associated merge distance.

\paragraph{Preprocessing.}
Before clustering, we remove redundant circles---specifically, any circle that completely contains another. This is performed by approximating each circle with its axis-aligned bounding box, sorting all boxes by decreasing radius, and using the R*-tree to query for candidate boxes that may fully cover the current one. We cap the total number of candidates examined across all queries at $O(n \log n)$, ensuring the preprocessing cost stays within budget at the expense of potentially missing some redundant circles. A theoretically exact $O(n \log n)$ solution exists via the Apollonius diagram (additively weighted Voronoi diagram) \parencite{Karavelas2002}, but we opt for the simpler spatial-index approach as missed redundancies have negligible impact on downstream clustering quality.

\paragraph{Merge Step.}
Let $c_1$ and $c_2$ denote the two circles chosen for merging. Their \emph{effective distance} is defined as
\[
d(c_1, c_2) = \|p_1 - p_2\| - r_1 - r_2,
\]
that is, the Euclidean gap between their boundaries. Intuitively, smaller values indicate strong overlap or proximity, which makes the pair more representative for early merging.

The merge procedure operates as follows:
\begin{enumerate}
    \item Select the pair $(c_1, c_2)$ with the minimal $d(c_1, c_2)$ value.
    \item Remove $c_1$ and $c_2$ from the R*-tree and from the corresponding neighbor structures.
    \item Compute their proxy circle $c'$.
    \item Insert $c'$ into the R*-tree and update all relevant neighbor lists.
    \item Insert $c'$ as an internal node in the binary tree, with $c_1$ and $c_2$ as its children.
\end{enumerate}

To mitigate alignment bias from axis-aligned boxes, we randomize the instance by applying a random global rotation to all circle centers. This randomization slightly improves expected merging quality. To maintain efficiency, we adopt the same general mechanism as \textcite{Formella2024}. Each active circle keeps a list of its current nearest neighbors, and a global min-heap stores tuples of the form $(d(c_i, c_j), i, j)$. When a pair is merged, only the affected neighbors of $c_i$ and $c_j$ are updated.

\paragraph{Approximate Neighbor Search.}
Because the effective distance $d(c_1, c_2) = \|p_1 - p_2\| - r_1 - r_2$ depends on both position and radius, exact nearest-neighbor search under this metric is impractical. Instead, we perform approximate search by querying the $k$ (small constant) closest bounding boxes (from a certain center $p_1$) and evaluating $d(c_1, c_2)$ explicitly for those candidates. This is inexact in two ways: boxes do not perfectly encapsulate the circles, and any value of $\|p_1 - p_2\| - r_2$ not greater than $0$ are treated the same. However, in practice, small constant values of $k$ suffice to preserve both runtime efficiency and clustering quality.

\begin{figure}[H]
    \centering

    \begin{subfigure}{0.32\textwidth}
        \centering
        \includegraphics[width=\textwidth]{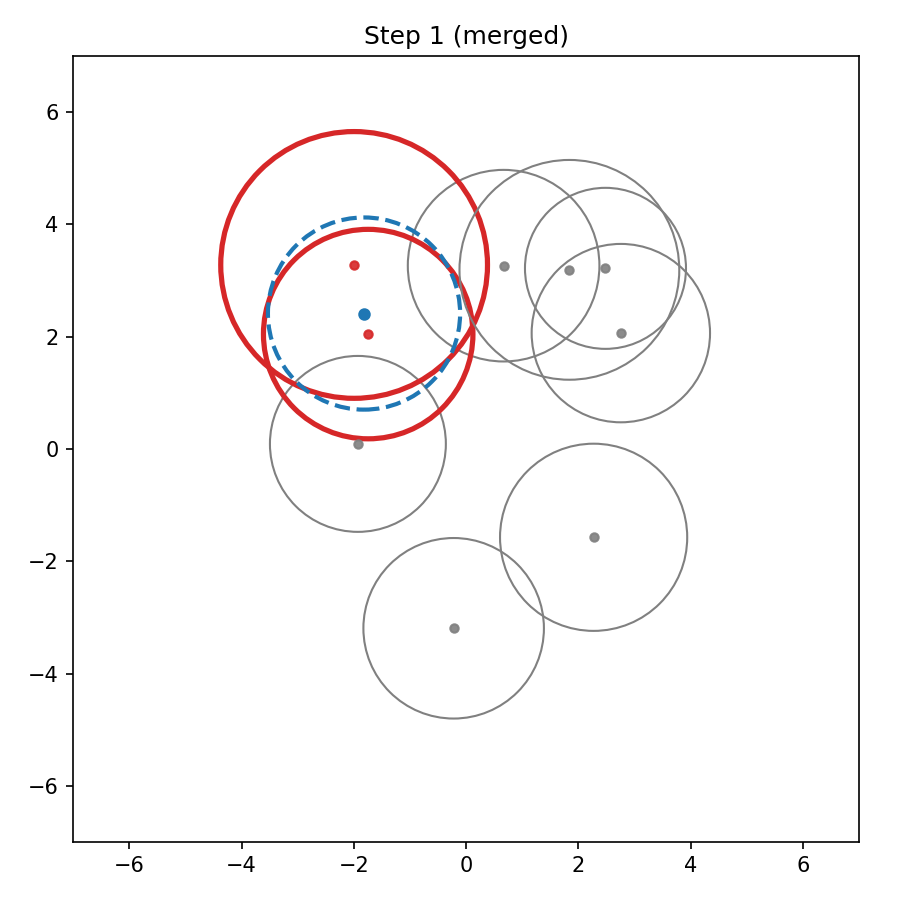}
        \caption{Step 1}
    \end{subfigure}
    \hfill
    \begin{subfigure}{0.32\textwidth}
        \centering
        \includegraphics[width=\textwidth]{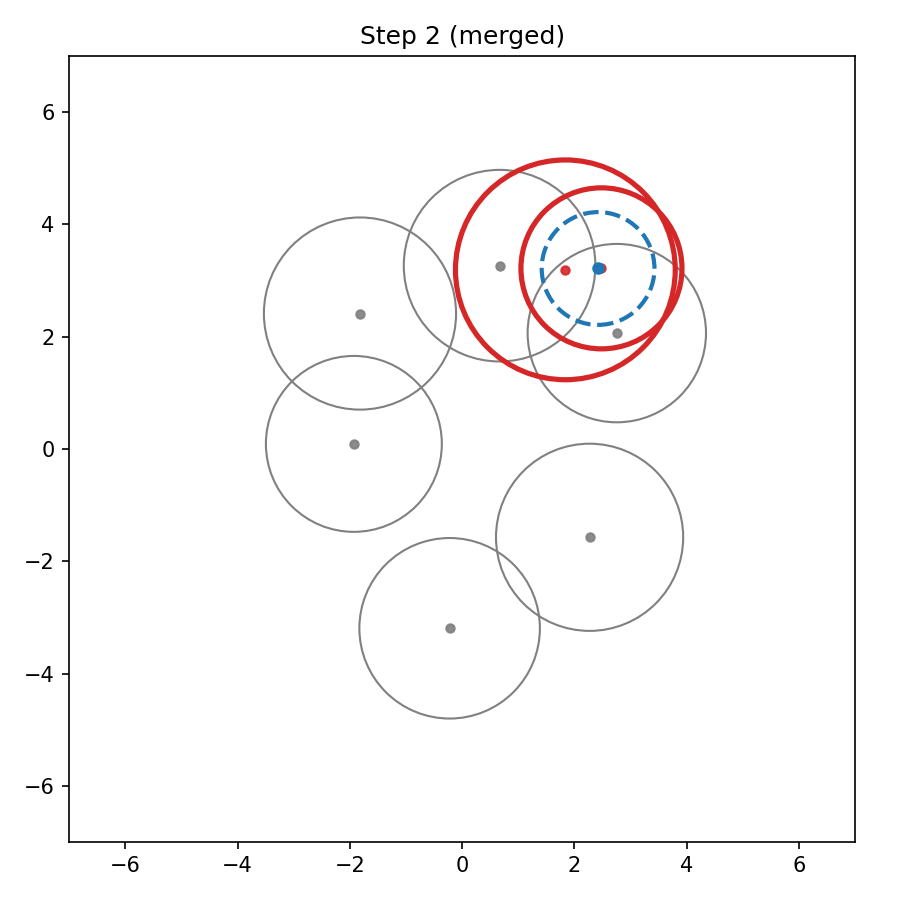}
        \caption{Step 2}
    \end{subfigure}
    \hfill
    \begin{subfigure}{0.32\textwidth}
        \centering
        \includegraphics[width=\textwidth]{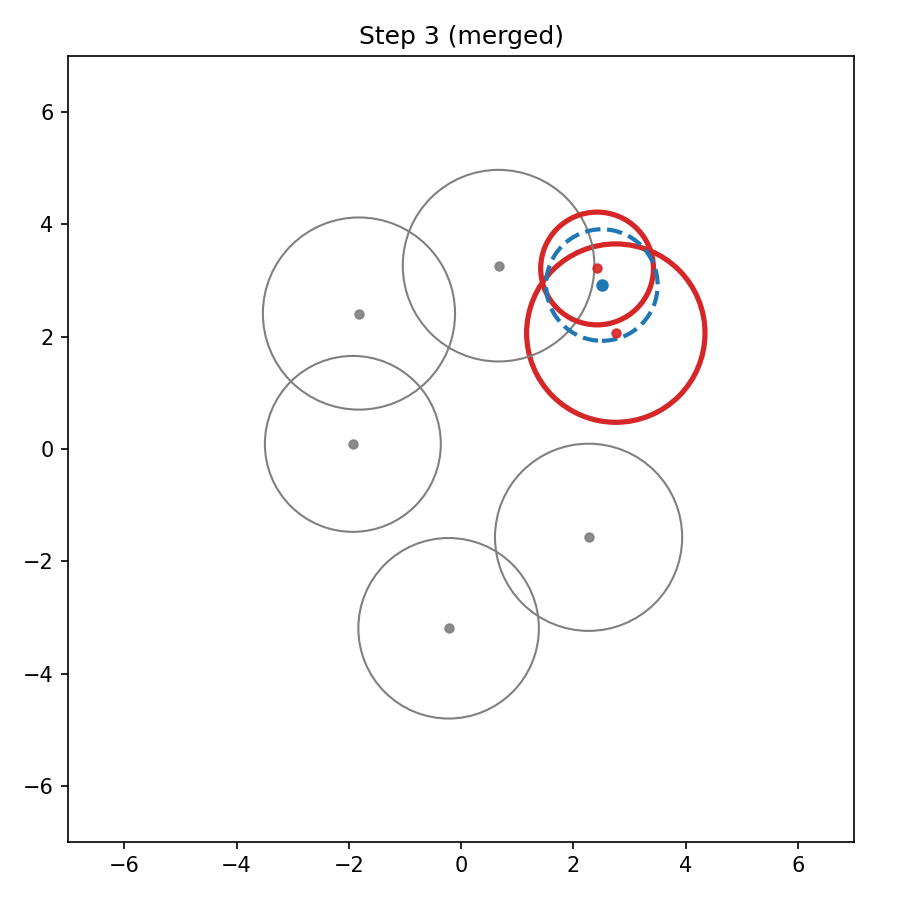}
        \caption{Step 3}
    \end{subfigure}

    \begin{subfigure}{0.32\textwidth}
        \centering
        \includegraphics[width=\textwidth]{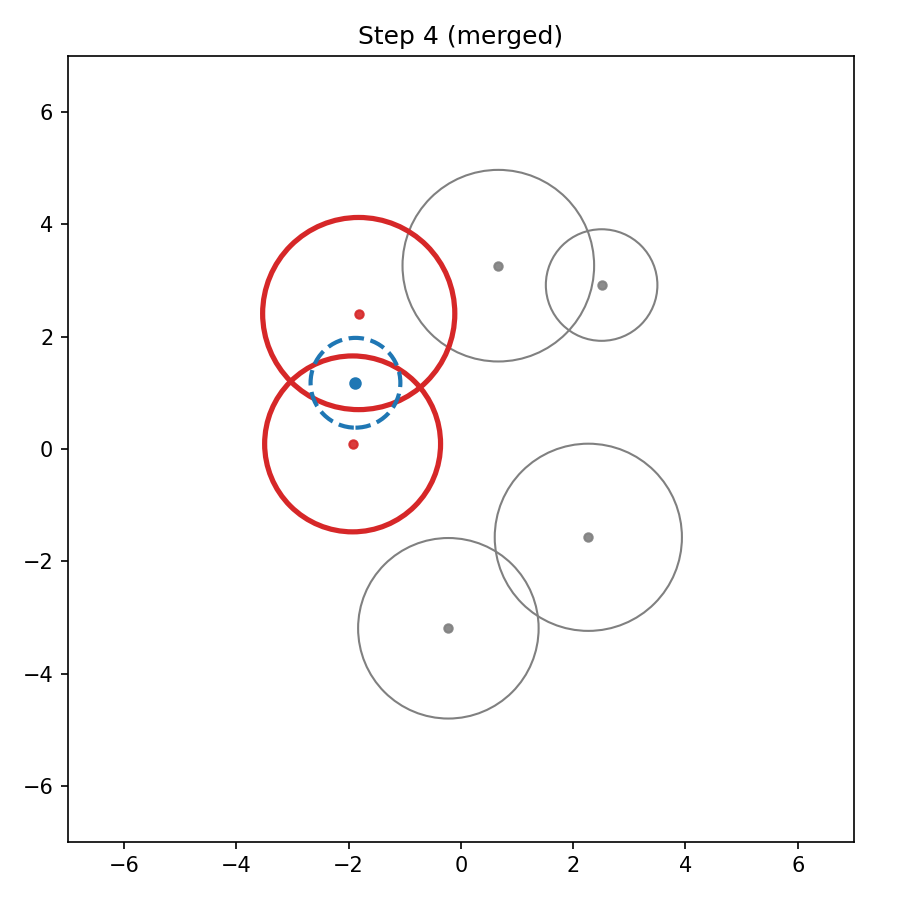}
        \caption{Step 4}
    \end{subfigure}
    \hfill
    \begin{subfigure}{0.32\textwidth}
        \centering
        \includegraphics[width=\textwidth]{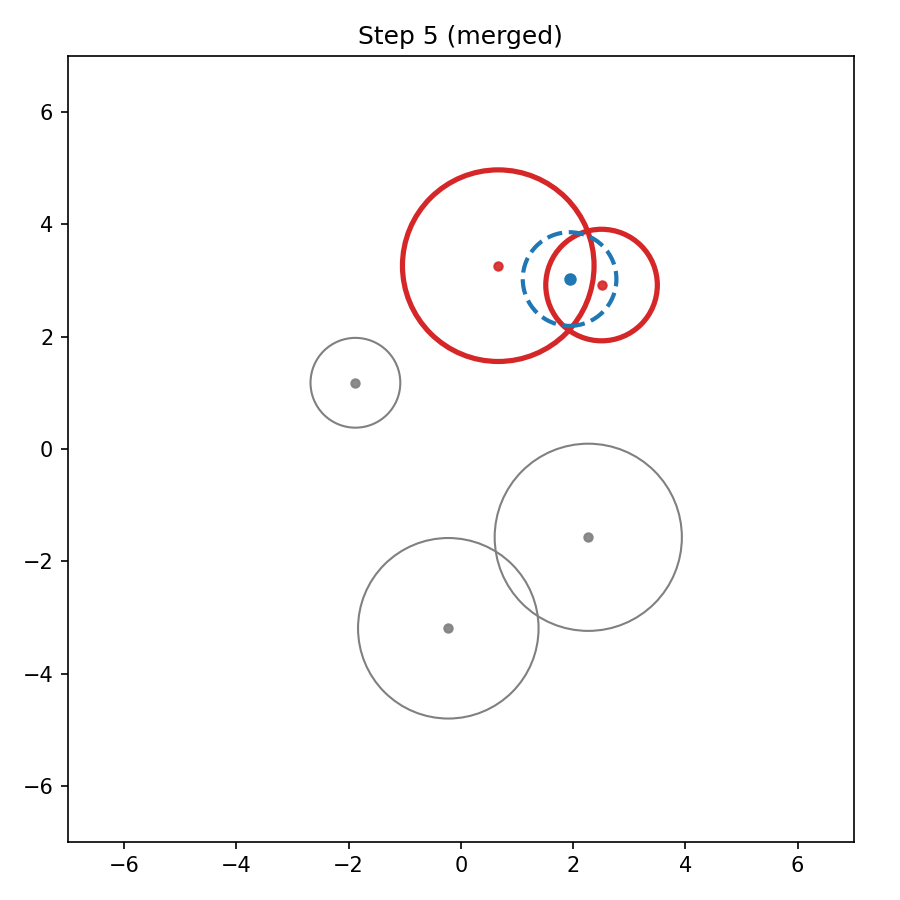}
        \caption{Step 5}
    \end{subfigure}
    \hfill
    \begin{subfigure}{0.32\textwidth}
        \centering
        \includegraphics[width=\textwidth]{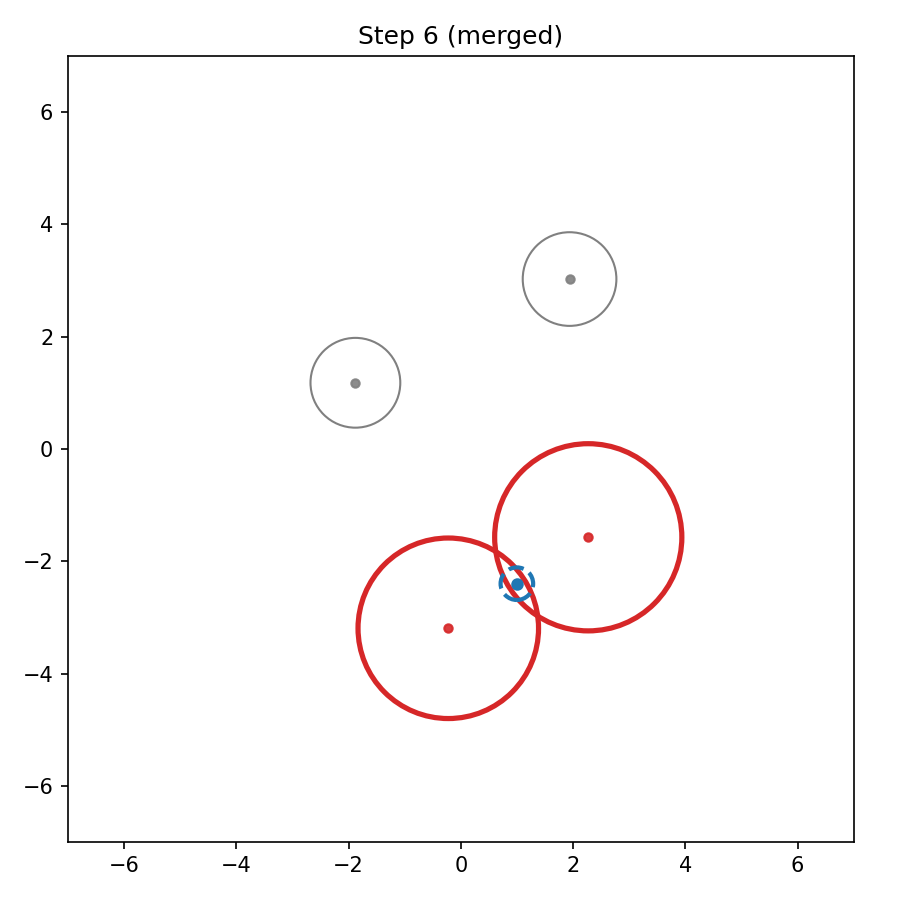}
        \caption{Step 6}
    \end{subfigure}

    \begin{subfigure}{0.32\textwidth}
        \centering
        \includegraphics[width=\textwidth]{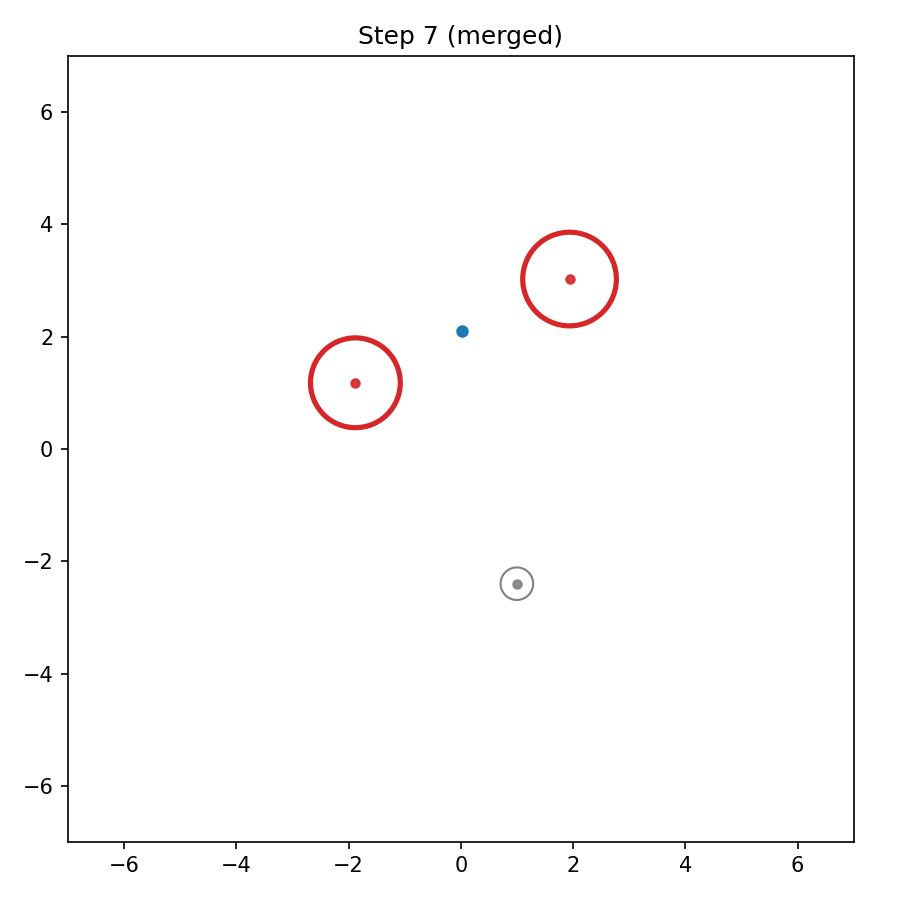}
        \caption{Step 7}
    \end{subfigure} 
    \hfill
    \begin{subfigure}{0.32\textwidth}
        \centering
        \includegraphics[width=\textwidth]{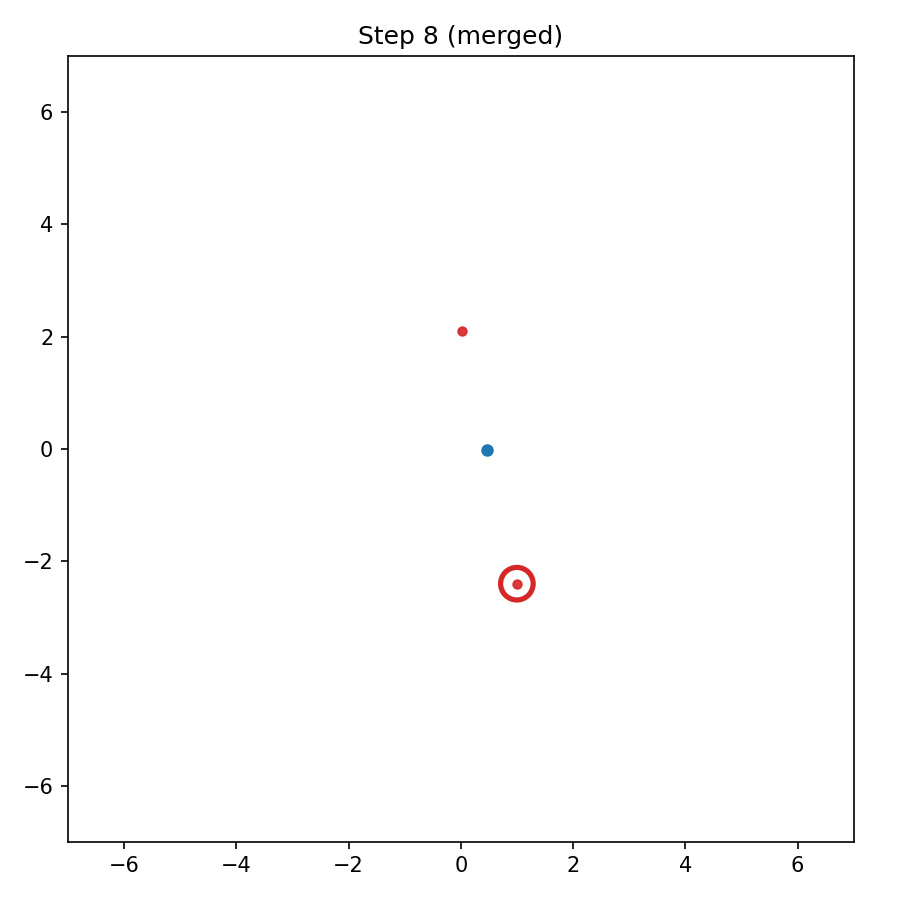}
        \caption{Step 8}
    \end{subfigure}
    \hfill
    \begin{subfigure}{0.32\textwidth}
        \centering
        \includegraphics[width=\textwidth]{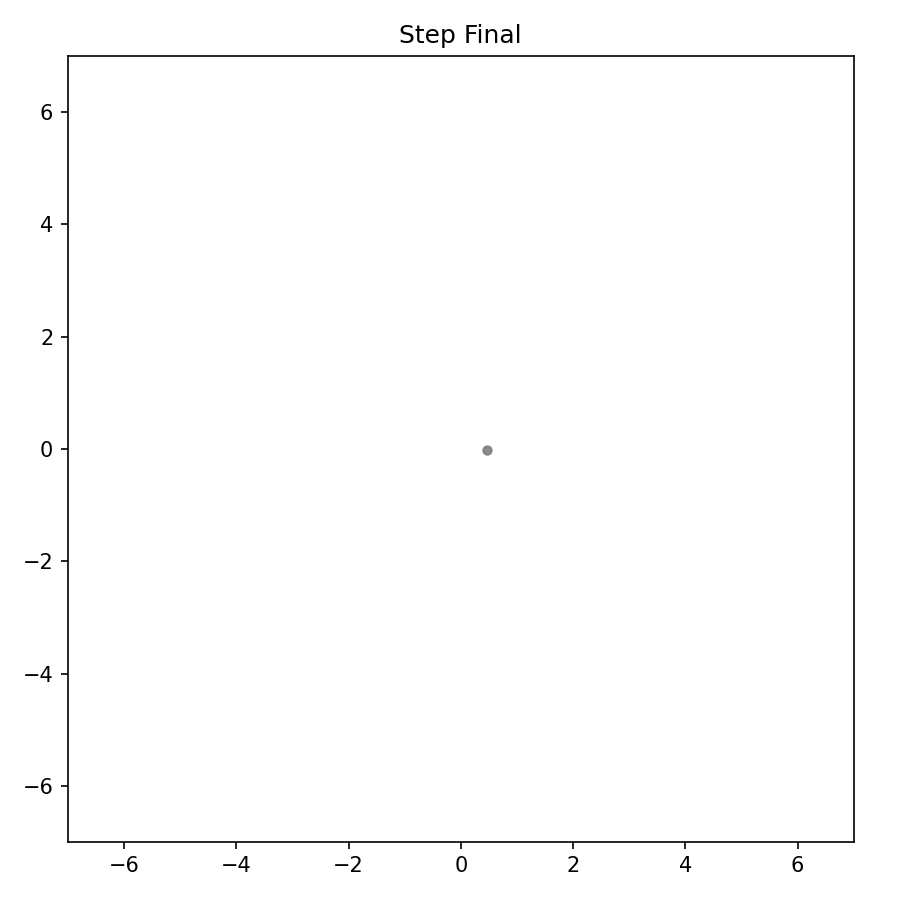}
        \caption{Final}
    \end{subfigure}

    \caption{Visualization of the clustering phase for a sample instance with $n=9$ disks. 
    Each subfigure shows one merge step, where the two closest circles (highlighted) are replaced by their proxy circle (dashed blue outline). 
    The final frame (bottom right) shows the single remaining cluster.}
    \label{fig:clustering_phase}
\end{figure}

\paragraph{Proxy Circle Computation.}
When merging two circles $c_1 = (p_1, r_1)$ and $c_2 = (p_2, r_2)$, we define their \emph{proxy circle} $c' = (p', r')$ as a compact approximation of the lens-shaped region covered by their union. In contrast to simply using the midpoint between centers, our implementation constructs $c'$ geometrically along the line connecting $p_1$ and $p_2$. 

Specifically, let $d = \|p_1 - p_2\|$ denote the center distance. If a circle covers another circle, we simply take the smaller circle. If the circles do not intersect at all ($d \geq r_1 + r_2$), we take the midpoint between the circle boundaries and radius $0$ (i.e. a single point). 

Otherwise, we construct a proxy circle. We determine the two points where the line connecting $p_1$ and $p_2$ intersects the boundaries of the two circles, and place the new center $p'$ at the midpoint between these boundary points. The resulting circle approximates the geometric ``lens'' formed by the intersection. We compute both the overlap depth
\[
\delta = \tfrac{1}{2}(r_1 + r_2 - d)
\]
and the half-chord length
\[
h = \sqrt{r_1^2 - a^2}, \quad a = \frac{r_1^2 - r_2^2 + d^2}{2d},
\]
which jointly define a plausible range for the new radius. To introduce mild stochastic diversity, the final radius $r'$ is drawn uniformly from the interval $[\delta, h]$. This randomization helps avoid degenerate clustering patterns and allows multiple independent runs to explore slightly different hierarchical structures.

Although this proxy construction is heuristic, it consistently yields stable and geometrically balanced merges in practice, while maintaining constant-time computation and preserving the algorithm's overall quasi-linear runtime.

\subsection{Construction Phase}
The construction phase of our algorithm follows the overall strategy of the pair-center algorithm described by \textcite{Formella2024}, with modifications to handle circular regions instead of point targets. The core objective is to insert each original circle into the tour by iteratively decomposing the hierarchical structure, maintaining a closed tour where a single tour point can represent multiple covered circles.

\paragraph{Tour Initialization and Expansion.}
The process begins by initializing the tour with a single tour point corresponding to the center point of the root of the binary tree. The root proxy circle itself is inserted into a global max-heap, which is ordered by the merge distance stored at the center node. The algorithm proceeds iteratively, expanding the subtrees until the max-heap is empty.

\paragraph{Circle Decomposition and Removal.}
In each step, the maximal entry in the heap---a circle $c = (O, r)$---is popped from the heap. This circle $c$ is then removed from the tour structure. This removal is handled conditionally: if the tour point that contained $c$ represented only this single circle, the tour point is entirely removed from the sequence. If the tour point represented a list of multiple circles (including $c$), only $c$ is removed from that list, and the tour point remains active.

\paragraph{Insertion of Generating Circles.}
The process then focuses on inserting the two generating circles, $c_i$ and $c_j$, that formed the proxy circle $c$. Insertion is performed by checking two possibilities:
\begin{itemize}
    \item \textbf{Check existing points:} The first priority is to assign the circle to an existing tour point to incur zero additional travel cost. This is possible if the existing tour point is not greater than the circle's radius. An R\*-tree of tour points is used for efficient querying of suitable existing points.
    \item \textbf{Create new tour point:} If no existing point can cover the circle, a new tour point $P$ must be created inside the circle $(O, r)$ and inserted into an existing tour segment $AB$. The optimal placement of $P$ must minimize the added tour length, $\Delta L = |AP| + |BP| - |AB|$.
\end{itemize}

\begin{figure}[H]
    \centering

    \begin{subfigure}{0.24\textwidth}
        \centering
        \includegraphics[width=\textwidth]{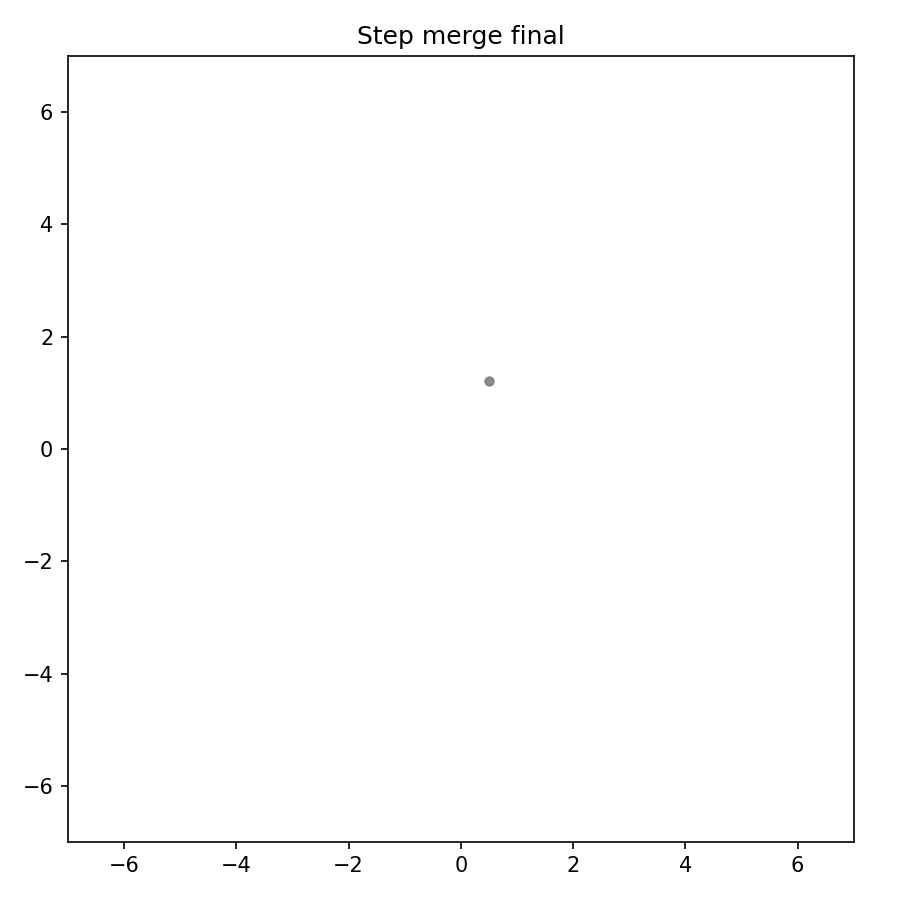}
        \caption{Initial}
    \end{subfigure}
    \hfill
    \begin{subfigure}{0.24\textwidth}
        \centering
        \includegraphics[width=\textwidth]{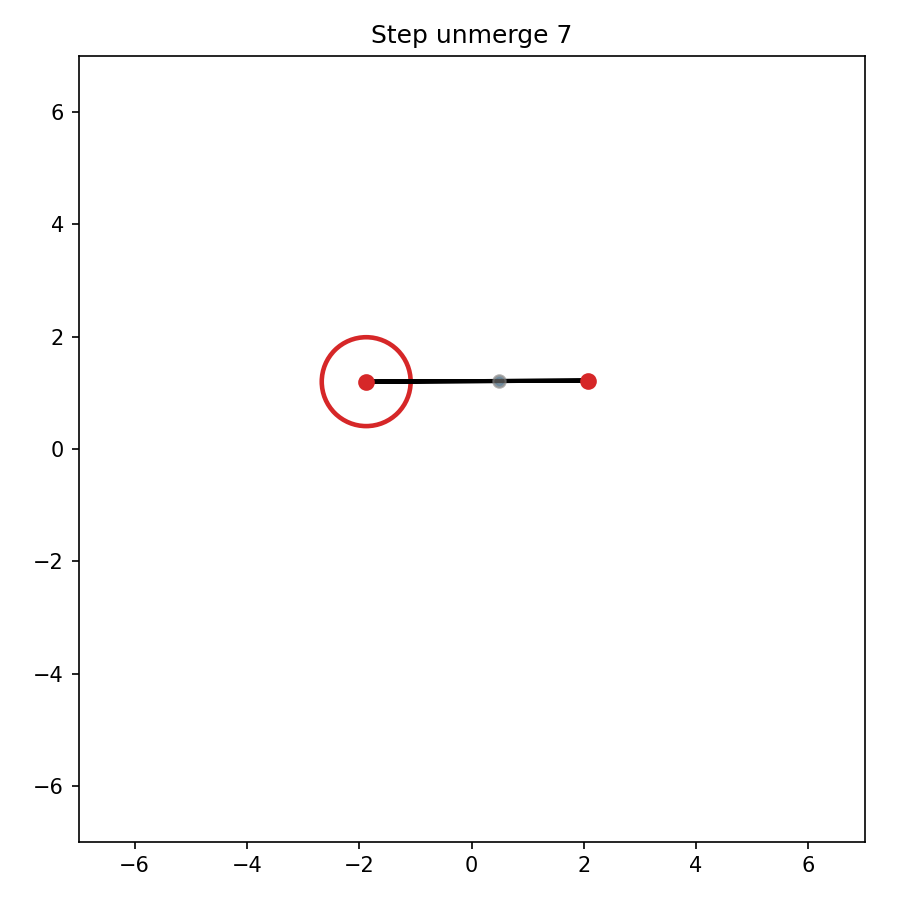}
        \caption{Step 1}
    \end{subfigure}
    \hfill
    \begin{subfigure}{0.24\textwidth}
        \centering
        \includegraphics[width=\textwidth]{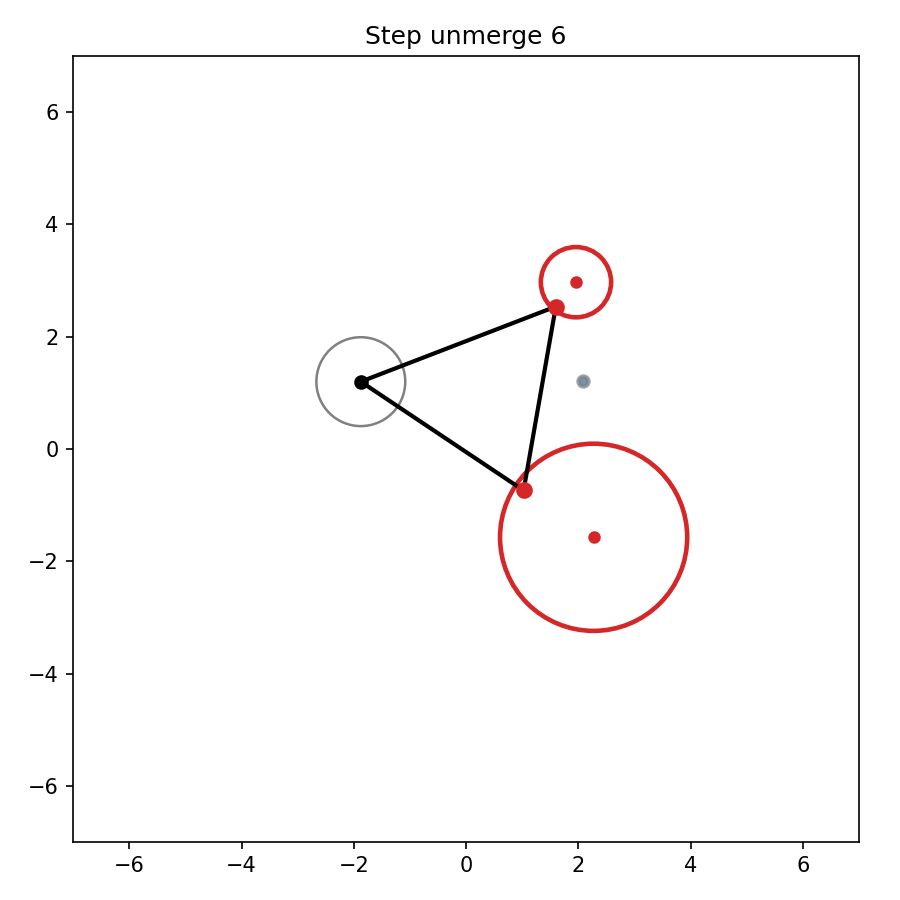}
        \caption{Step 2}
    \end{subfigure}
    \hfill
    \begin{subfigure}{0.24\textwidth}
        \centering
        \includegraphics[width=\textwidth]{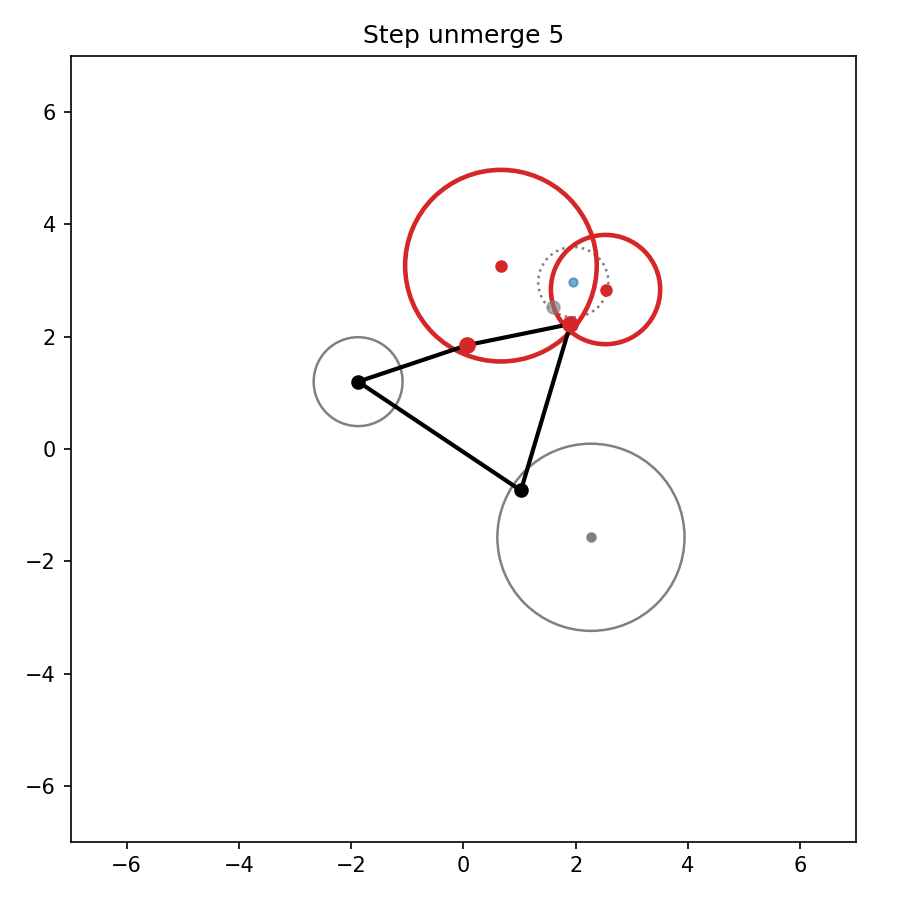}
        \caption{Step 3}
    \end{subfigure}

    \vspace{0.6em}

    \begin{subfigure}{0.24\textwidth}
        \centering
        \includegraphics[width=\textwidth]{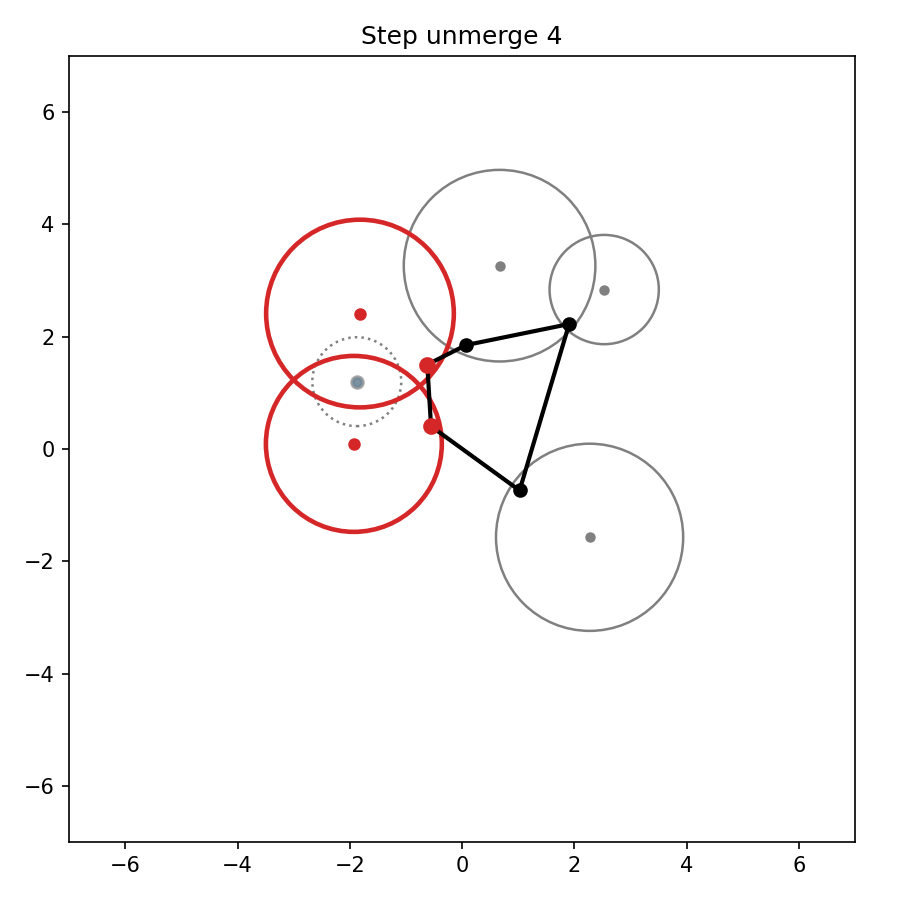}
        \caption{Step 4}
    \end{subfigure}
    \hfill
    \begin{subfigure}{0.24\textwidth}
        \centering
        \includegraphics[width=\textwidth]{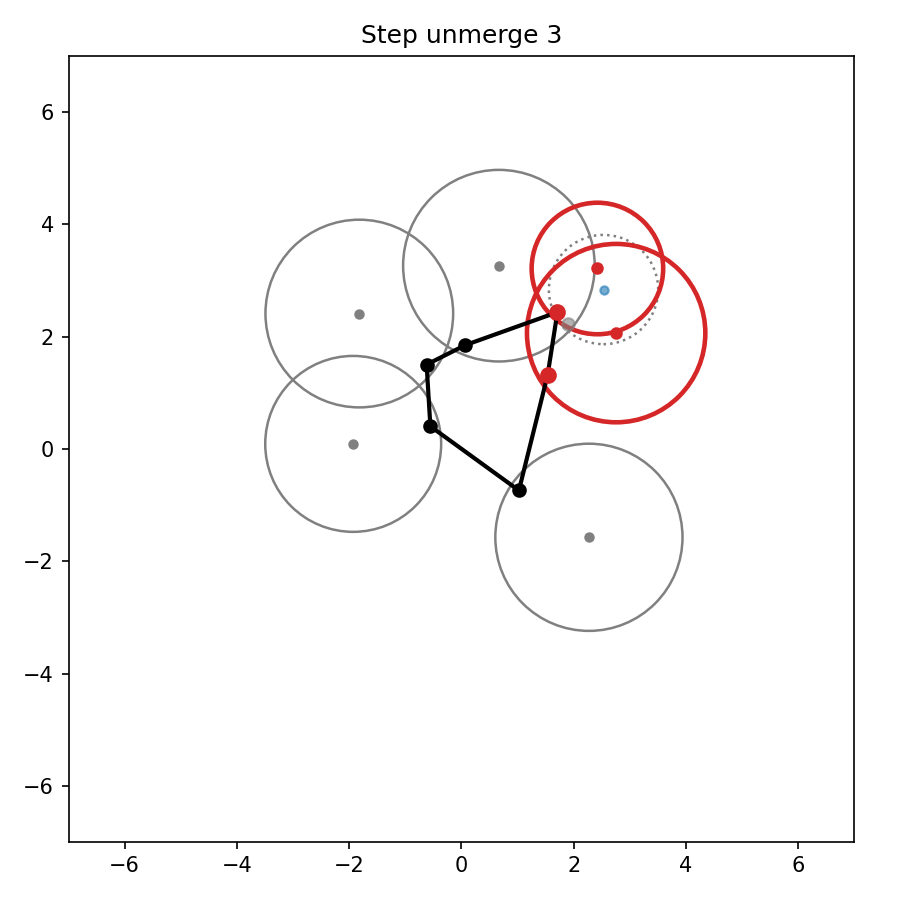}
        \caption{Step 5}
    \end{subfigure}
    \hfill
    \begin{subfigure}{0.24\textwidth}
        \centering
        \includegraphics[width=\textwidth]{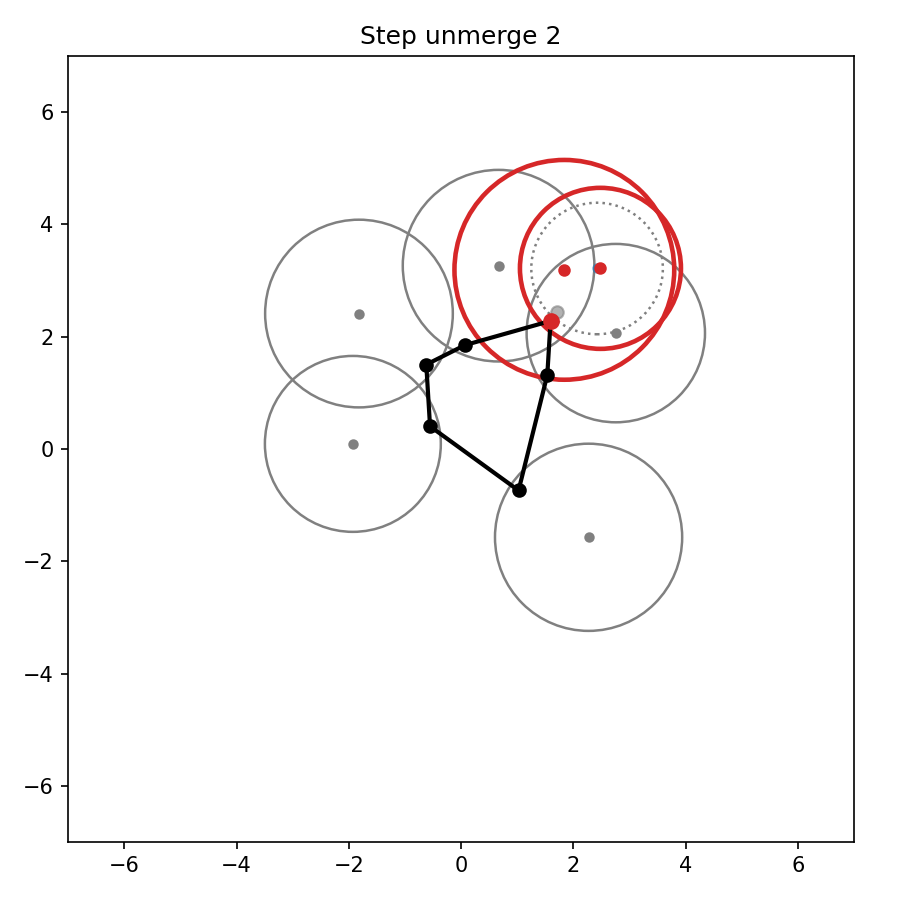}
        \caption{Step 6}
    \end{subfigure}
    \hfill
    \begin{subfigure}{0.24\textwidth}
        \centering
        \includegraphics[width=\textwidth]{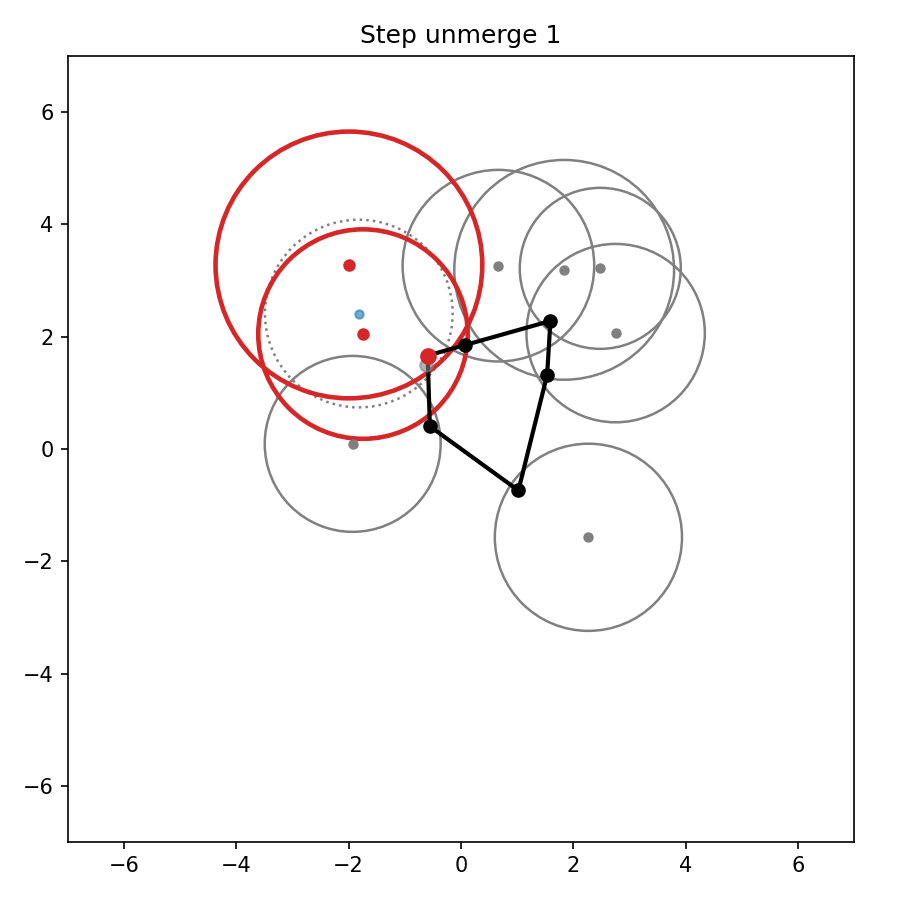}
        \caption{Step 7}
    \end{subfigure}

    \caption{Visualization of the construction phase for the same sample instance.
    The process starts from the final proxy circle obtained in the clustering phase (top left) and iteratively expands proxy circles into their generating circles while updating the tour.
    Newly inserted circles and tour points are highlighted, while removed proxy elements are shown in a faded, dashed style.
    Note that we do not show point optimizations here.}
    \label{fig:unmerge_phase}
\end{figure}

\paragraph{Approximate Tour Placement.}
Finding the globally optimal insertion point is computationally intensive (this is Alhazen's problem). Therefore, we use an approximation: we first select the $k$ nearest tour segments $AB$ using an R\*-tree of tour segments. For each segment, we put a provisional new point $P$ at the bisection of the angle $\angle AOB$, where $O$ is the circle center. We then compare all points $P$ to find the best-fitting segment. This method is a fast and effective approximation of the optimal point \parencite{LeiHao2024}.  Optionally, we can also apply a single Newton-Raphson iteration to refine this approximate placement. If the generating circles $c_i$ or $c_j$ are themselves proxy circles (internal nodes), they are inserted into the max-heap for further hierarchical expansion. The algorithm terminates once the max-heap is empty, indicating that all original leaf-node circles have been assigned to tour points, resulting in a complete tour.

\begin{figure}[H]
    \centering
    \includegraphics[width=0.6\linewidth]{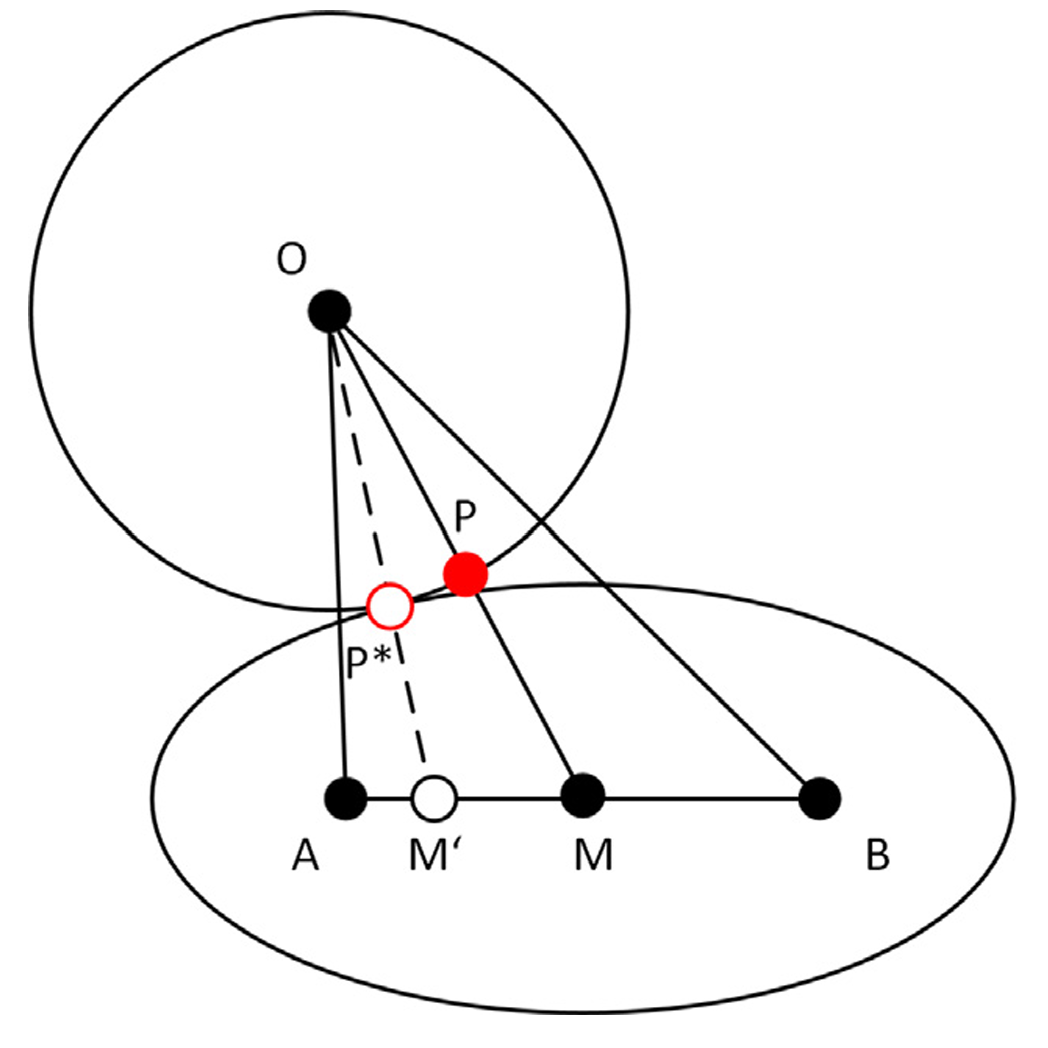}
    \caption{Approximate solution to the Alhazen problem for placing a new tour point within a circle. The bisection of angle $\angle AOB$ is used as a fast approximation of the optimal insertion point \parencite{LeiHao2024}.}
    \label{fig:alhazen}
\end{figure}

\subsection{Point Optimization}

During our construction, some tour points may be geometrically and structurally suboptimal. To further improve solution quality without sacrificing runtime scalability, we introduce two lightweight local optimization mechanisms: \emph{reinsertion} and \emph{point reoptimization}. Both are designed to be incremental and self-regulating, preserving the algorithm's overall quasi-linear behavior while opportunistically improving the tour.

\paragraph{Reinsertion.}
During construction, tour points are created incrementally and may become suboptimal as the tour evolves. To correct this, we periodically remove a tour point and reinsert all circles associated with it as if they were new circles. This procedure allows local restructuring of the tour without recomputing the full solution.

Because global reinsertion is computationally expensive, we perform it selectively based on a heuristic “energy” model. Each tour point maintains an energy value $E$ that quantifies its local instability. Whenever a new circle is inserted into a tour point, we add $+3$ energy to that point and $-1$ from each of its immediate neighbors. When the energy of a tour point reaches zero, it triggers a reinsertion event: the point is removed from the tour, and all circles it represented are reinserted individually following the same rules described in the construction phase.

This mechanism adaptively prioritizes reinsertion in regions of the tour where structural changes are concentrated, while naturally suppressing redundant updates in stable areas. Our policy also limits the number of reinsertion events so that we can preserve the near-linear scaling of the algorithm.

\paragraph{Point Reoptimization.}
A second optimization addresses the geometric placement of existing tour points. Each point is originally positioned based on the local approximation used at its first insertion, which may no longer be optimal once additional circles are added or neighboring segments shift.

The reoptimization problem is as follows. Given a tour segment $(A,B)$ and a tour point $P$ representing a set of circles $\{(O_i, r_i)\}$, we wish to find a new position $P^\star$ within the feasible intersection region of these circles that minimizes
\[
\Delta L = |A P^\star| + |B P^\star| - |A B|.
\]
This optimization is nontrivial even for a single circle, and becomes analytically intractable for multiple overlapping disks. We therefore adopt a fast geometric approximation that suffices for local improvement.

\begin{enumerate}
    \item If the segment $AB$ intersects the common feasible region of all circles, the optimal solution incurs zero additional cost; in this case, we simply take a point in the intersection.
    \item Otherwise, the optimal $P^\star$ must lie on the boundary of the intersection region. We approximate this by performing a single gradient-descent step: we compute the gradient of $\Delta L$ with respect to $P$ and move $P$ maximally along that direction until it would leave any of the circles. This yields a fast, stable adjustment that effectively reduces local tour length.
\end{enumerate}

To control runtime, reoptimization is triggered according to a simple exponential schedule. Specifically, a tour point is reoptimized whenever the number of insertions of circles corresponding to that point reaches $2^n$ for some integer $n$. This ensures that frequently updated regions receive more attention while overall computational effort remains bounded.

\begin{figure}[H]
    \centering
    \includegraphics[width=0.6\linewidth]{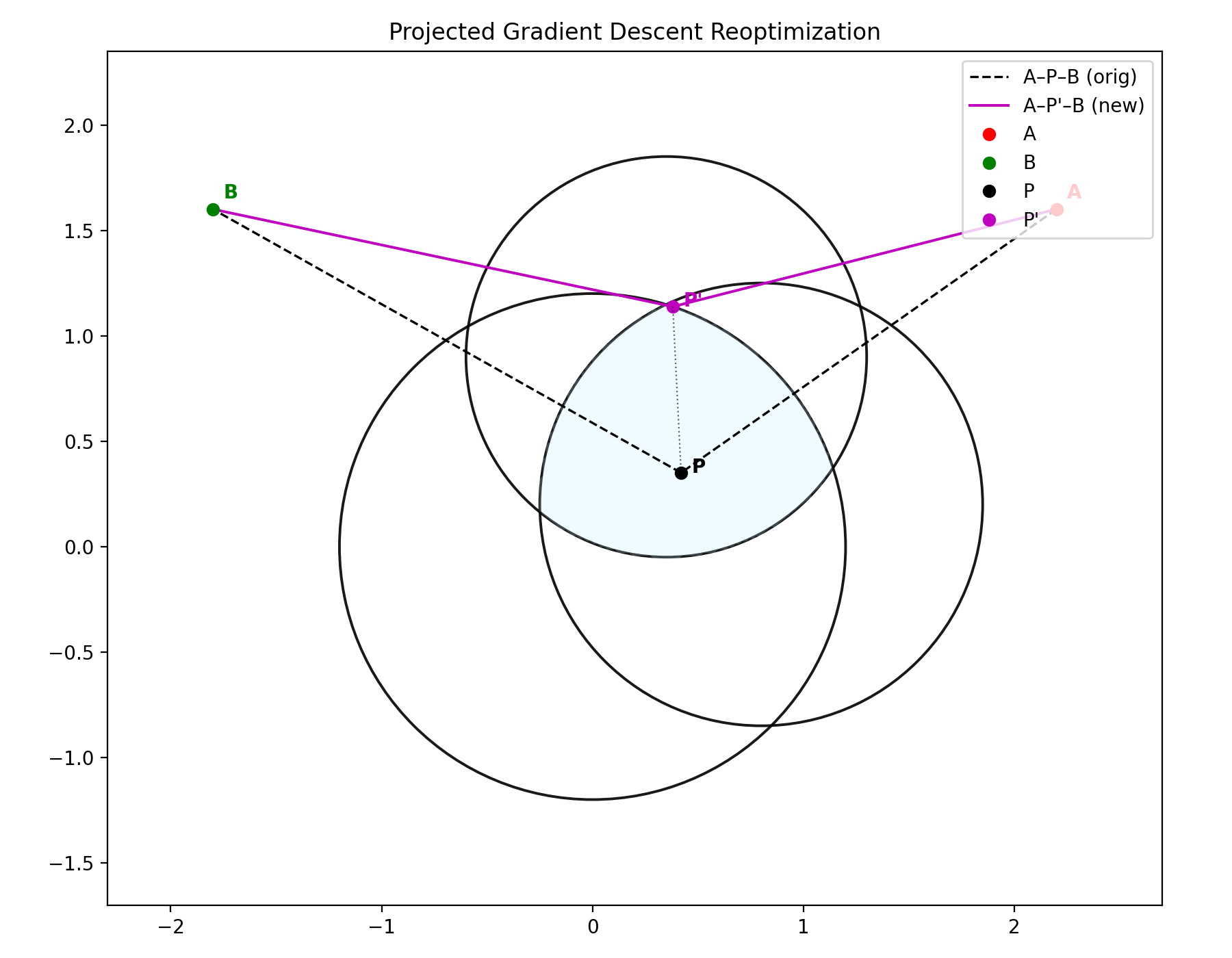}
    \caption{Illustration of the point reoptimization step. The new position $P^\star$ is obtained by projecting the current point $P$ toward the gradient direction of decreasing tour cost, constrained to remain within all corresponding circles.}
    \label{fig:reoptimize}
\end{figure}

Together, these two mechanisms maintain geometric consistency of the tour and correct local inefficiencies introduced during hierarchical construction. Both procedures are incremental, spatially localized, and amortized over the construction process, maintaining the total runtime while providing measurable quality improvements.

\subsection{Runtime Analysis}

We analyze the expected running time of the algorithm.  As in the clustering phase, we measure time in terms of basic geometric operations (R\*-tree queries, insertions, deletions, and heap operations). We assume the standard randomized/average-case performance bounds for R\*-trees (or similar balanced spatial search structures): each query, insertion or deletion takes expected $O(\log n)$ time. All claims below are in expectation with respect to the randomization used in preprocessing and the probabilistic behavior of the spatial index.

\paragraph{Single-query cost.}
Retrieving the $k$ nearest bounding boxes (or segments or points) from the R\*-tree costs expected $O(\log n + k)$.  We repeatedly use this fact to bound the cost of approximate nearest-neighbor and segment queries.

\subsubsection{Preprocessing phase}
We first sort the $n$ axis-aligned bounding boxes by decreasing radius. This costs $O(n \log n)$.

After sorting, we process each box once: for the current box we query the R\*-tree for candidate boxes that may fully contain it, and test those candidates for containment. Let $k_i$ denote the number of candidates examined for the $i$-th query. The total query cost is
$$\sum_{i=1}^n \bigl(O(\log n) + k_i\bigr) = O(n\log n) + \sum_i k_i.$$
Bounding $\sum_i k_i$ is non-trivial: in the average case $k_i$ is constant, but a bounding box may generate many false-positive hits from the spatial index in the worst case. We therefore cap $\sum_i k_i$ at $O(n \log n)$ explicitly, terminating further candidate examination once this budget is exhausted. Combining terms gives
$$T_{\text{preproc}} = O(n\log n) + O(n\log n) = O(n\log n).$$

\subsubsection{Clustering phase}
The clustering phase performs $n-1$ merge iterations.  In each iteration we:
\begin{itemize}
  \item extract the current closest pair from a min-heap (expected $O(\log n)$),
  \item remove two entries from the R\*-tree (expected $O(\log n)$ each),
  \item compute a proxy circle (constant-time geometric work),
  \item insert the proxy into the R\*-tree (expected $O(\log n)$), and
  \item update a constant number of neighbor lists (each update involves at most a constant number of R\*-tree / heap operations and therefore costs expected $O(\log n)$).
\end{itemize}
Summing the dominant $O(\log n)$ costs over the $O(n)$ merges yields
\[
T_{\text{cluster}} = O(n \log n).
\]

\subsubsection{Construction phase (excluding point-optimizations)}
During construction we perform $O(n)$ heap pops and insertions (each costing expected $O(\log n)$), and for each insertion we perform a bounded number of R\*-tree queries or updates (nearest-point checks, segment queries, and structure updates), each costing expected $O(\log n)$.  Thus the dominant cost of construction (excluding point-optimizations) is
\[
T_{\text{construct}} = O(n \log n).
\]

\subsubsection{Cost of reinserts}
During the construction phase, we also use the point reinsertion mechanism described above. Each circle reinserted costs $O(\log n)$ expected time, as described in the construction phase. To bound the runtime of our reinsertion policy, we must bound the total number of reinsertions performed. Note that this bound depends on the number of \emph{circles} that are reinserted, not the number of \emph{tour points} that are reinserted.

We prove this using a \emph{gadget problem}. In this gadget problem, there are no circle deletions and energy subtractions can go into any node, not just neighbors. This also allows us to discount the tour structure entirely, and allow our process to work on a infinite set of abstract nodes. Note that these changes can only increase the number of reinsertions performed, so a bound on the gadget problem also bounds our actual reinsertion cost.

\paragraph{Gadget problem description.}
We consider a collection of nodes (finite or infinite in number), each of which is associated with two integer-valued parameters:
\[
E(v) \in \mathbb{Z}_{\ge 0} \quad \text{(energy)}, \qquad
W(v) \in \mathbb{Z}_{\ge 0} \quad \text{(weight)}.
\]
Initially, all nodes satisfy $E(v) = W(v) = 0$.

We define an \emph{operation} as the following sequence of updates:
\begin{enumerate}
    \item \emph{Insertion.} Select a node $v_+$ and perform the update
    \[
    E(v_+) \mathrel{+}= 3, \qquad W(v_+) \mathrel{+}= 1.
    \]
    \item \emph{Energy reduction.} Then, twice, select (possibly different) nodes $v_-$ and perform the update
    \[
    E(v_-) \mathrel{-}= 1.
    \]
\end{enumerate}

Whenever a node $v$ reaches energy $E(v) = 0$, we say that $v$ \emph{resets}.  
Let $w = W(v)$ be its weight at the moment of reset.  
Upon reset, we perform:
\[
W(v) \gets 0,
\]
and immediately execute $w$ new operations as defined above.
These induced operations are called \emph{extra operations}.

More concretely, imagine a stack of updates. Initially, we push $3n$ updates onto the stack, alternating between two energy reductions and one insertion. We then repeatedly pop updates from the stack and execute them. Whenever an energy reduction causes a node to reset, we push $3w$ new operations onto the stack in the same manner, where $w$ is the weight of the node at reset time. The total number of operations is equal to the number of insertions performed.

Suppose we begin with $n$ \emph{original operations} (executed according to the above rule, including all recursively spawned extra operations). Let $X$ denote the total number of extra operations generated in the entire process. Show that the process always terminates (i.e. $X$ is finite) and prove that $X \in O(n)$.

\paragraph{Proof of gadget problem bound.}
Start with $n$ initial operations. Let $X$ be the total number of extra operations ever spawned. Then the process terminates and
\[
X \le 2n.
\]
We proceed by induction on the number of initial operations performed.

\emph{Base case: $k = 1$.}
Execute the first initial operation: it does one $(+3)$ (on some node) and two $(-1)$ updates. No node can reach energy $0$ as a result of this single operation, because the only possible decrements are the two $(-1)$s produced by this operation (they can reduce the node that received $(+3)$ by at most $2$, leaving energy $\ge 1$). Hence no extra operation can be spawned. So after the first initial op finishes (including any extras it might spawn---there are none) we have a finite prefix, and the number of extra ops produced so far $X_1 = 0 \le 2\cdot 1$.

\emph{Inductive hypothesis.}
Fix $k\ge 1$. Assume that after performing the first $k$ initial operations (together with all extra ops they spawn) the process finishes a finite prefix and the total number of extras so far satisfies
\[
X_k \le 2k.
\]

\emph{Inductive step.}
Suppose for contradiction that after performing the first $k+1$ initial operations (and following all cascading extra ops) either (a) the process does not terminate, or (b) it terminates but the total number of extra ops $X_{k+1}$ produced by the first $k+1$ initial ops satisfies $X_{k+1}>2(k+1)$.

If (a) holds, then the number of extra ops produced is infinite; in particular there is a finite time at which the count of extra ops exceeds $2(k+1)$. Thus in both (a) and (b) we can find a finite prefix of the run that contains the first $k+1$ initial operations and whose number of extra ops $X'$ satisfies
\[
X' > 2(k+1).
\]
Choose the earliest (i.e., shortest) such finite prefix---earliest in the sense of the first time during the run the extra-op count becomes $>2(k+1)$. By construction this prefix is finite and contains exactly $(k+1)+X'$ executed operations (initial + extra).

Now count updates in that prefix:

\begin{itemize}
\item Every executed operation contributes one $(+3)$ and two $(-1)$ updates. So the prefix contains total $(+3)$-events equal to $(k+1)+X'$ and total $(-1)$-events equal to $2((k+1)+X')$.
\item Each extra operation in the prefix corresponds to a reset that consumed a weight unit which in turn came from a prior $(+3)$. To produce one extra operation (i.e. to consume one weight-unit) the corresponding $(+3)$ must have been fully drained by three $(-1)$ decrements on that node. Thus the decrements that were used to create these $X'$ extra ops are at least $3X'$ in number.
\end{itemize}

But these decrements must be among the $2\big((k+1)+X'\big)$ total decrements that actually occurred in the prefix. Therefore
\[
3X' \le 2\big((k+1)+X'\big).
\]
Rearranging:
\[
3X' \le 2k+2 + 2X' \quad\Longrightarrow\quad X' \le 2(k+1).
\]
This contradicts $X' > 2(k+1)$, so our assumption was false.

Therefore neither (a) nor (b) can happen: the run started by the first $k+1$ initial operations (together with all extras they spawn) must terminate and its number of extra ops satisfies $X_{k+1}\le 2(k+1)$. This completes the inductive step.

By induction, for every $k\le n$ the prefix that contains the first $k$ initial operations (and all extras they spawn) is finite and the number of extra operations produced satisfies $X_k\le 2k$. Applying this to $k=n$ gives termination for the whole process and the desired bound
\[
X \le 2n,
\]
so $X=O(n)$.

\subsubsection{Cost of point reoptimizations}
A reoptimization for a tour point operates only on the (usually small) list of circles associated to that point. By our exponential scheduling (reoptimize after $1,2,4,\dots$ insertions), the total amount of work spent on reoptimizing a single point over the entire run is $O(s)$ where $s$ is the total number of insertions into that point; summing over all points yields an overall reoptimization cost that is proportional to the total number of insertions, i.e. $O(I) = O(n)$. Thus
\[
T_{\text{reopt}} = O(n).
\]

\subsubsection{Total expected time}
Combining the dominant terms,
\[
T_{\text{total}} \;=\; O(n \log n).
\]

\paragraph{Remarks and caveats.} This analysis depends on the logarithmic performance of the spatial index that we use. In theory, these indexes can degrade to worse behavior in bad cases, so the actual worst-case runtime is higher and depends on the spatial index data structure used. In practice, however, the expected $O(n \log n)$ behavior or better is observed.

\section{Initializing MA-CETSP with Constructed Tours}

The proposed constructor is fast and randomized, allowing it to generate many high-quality tours with different visit orders at low cost. We use these tours to initialize MA-CETSP \parencite{LeiHao2024} in place of its native population generation procedure.

A constructed tour may use one visit point to cover several disks, whereas MA-CETSP assigns a separate visit to every disk. We therefore expand each shared visit into one point per covered disk without changing the polygonal path or its length. We retain a population of short tours with sufficiently different visit orders and pass them to MA-CETSP. During initialization, MA-CETSP applies its usual LKH ordering improvement and SOCP point optimization to each supplied tour before admitting it to the population. The remainder of the MA-CETSP search is unchanged.

\section{Benchmark Evaluation and Comparison}

\subsection{Benchmark Instances}

We evaluate our algorithm on the \textcite{Mennell2009} dataset; see \textcite{LeiHao2024} for an overview of its composition and benchmark usage in the CETSP literature.

These instances were originally obtained from three sources---\textbf{TSPLIB} \parencite{TSPLIB1991}, \textbf{Teams}, and \textbf{Geometric}---and consist of a total of 62 benchmark cases:
\begin{itemize}
  \item \emph{TSPLIB subset.} 28 instances generated from seven graphs (\texttt{d493}, \texttt{dsj1000}, \texttt{kroD100}, \texttt{lin318}, \texttt{pcb442}, \texttt{rat195}, and \texttt{rd400}) with sizes ranging from 100 to 1000 nodes. These include two with uniformly generated nodes (\texttt{d493}, \texttt{dsj1000}), two with clustered nodes (\texttt{kroD100}, \texttt{lin318}), two drill-press problems (\texttt{pcb442}, \texttt{rat195}), and one with nodes arranged roughly along lines (\texttt{rd400}).
  \item \emph{Teams subset.} 14 instances derived from seven graphs containing the terms \emph{team} or \emph{bonus}, introduced by \textcite{Gulczynski2006}, comprising three uniformly distributed and four clustered point sets.
  \item \emph{Geometric subset.} 20 instances from \textcite{Mennell2009}, including 9 \emph{Bubbles} (overlapping concentric squares), 5 \emph{ConcentricCircles}, 5 \emph{RotatingDiamonds} (nonoverlapping concentric squares), and one \emph{chaoSingleDep} instance based on \textcite{Chao1993}.
\end{itemize}

Following \textcite{LeiHao2024}, the dataset can be grouped by overlap characteristics and disk radii:
\begin{itemize}
  \item \emph{Group 1 (G1).} 27 instances with different overlap ratios (OR), identical radii within each instance.
  \item \emph{Group 2 (G2).} 21 instances with fixed overlap ratios (0.02, 0.1, and 0.3), again identical radii within each instance.
  \item \emph{Group 3 (G3).} 14 instances with heterogeneous radii.
\end{itemize}

\subsection{Dataset Discrepancies and Reconstruction}

The \emph{Mennell dataset} currently distributed on public repositories differs substantially from the version used in prior studies.
In the current release,
\begin{itemize}
  \item the \emph{TSPLIB subset} no longer includes the identical-radius variants,
  \item several \emph{new instances} (the \texttt{nlpProb} series) appear, and
  \item a \emph{large-scale} 13509-circle test case (\texttt{usa13509.txt}) has been added.
\end{itemize}

This mismatch complicates reproducibility and benchmarking, as overlap ratios and radii are inconsistent across sources.
Furthermore, the calculation of the \emph{overlap ratio} (OR) is ambiguous in the literature:
\textcite{Mennell2009} and \textcite{DiPlacido2022} describe different procedures for computing OR, and the provided data does not align with either definition.

To maintain compatibility with prior CETSP works, we reconstructed the original identical-radius test cases.
Since the circle centers and solution tours are known, and the radii of the circles are equal, the radius can be recovered directly as
\begin{equation}
  r = \max_i \min_j \| p_i - c_j \|,
\end{equation}
where \(p_i\) are points on the optimal tour and \(c_j\) are the circle centers.
Using this method, we successfully recovered all identical-radius test instances corresponding to those used by \textcite{LeiHao2024} and earlier studies.

\subsection{Comparison Protocol}

All experiments were conducted on the reconstructed benchmark suite, ensuring consistency with previous publications. We report the standalone constructor and hybrid pipeline separately because they answer different quality--runtime questions.

\paragraph{Standalone protocol.}
The standalone results report the best of 1,000 independent construction runs, except for \texttt{bonus1000}, which uses 10,000. Table~\ref{tab:standalone-comparison} compares these tours with the best values reported by \textcite{LeiHao2024}. Table~\ref{tab:runtime_adjusted} contains only standalone runtimes.

For completeness, we also evaluate our method on the newer \texttt{nlpProb} and \texttt{usa13509} instances; however, since no prior results exist for these cases, it would be meaningless to include them in the main comparative analysis.\footnote{The results are still available in the dataset for future benchmarking.}

\paragraph{Hybrid protocol.}
Each instance receives 20 independent hybrid runs, matching the number of runs reported by \textcite{LeiHao2024}. Each run generates 1,000 construction tours (10,000 for \texttt{bonus1000}) and selects a population of 20. MA-CETSP uses a 5,000-generation limit, population size 20, fitness weight 0.96, edit-distance threshold 5, neighbor list size 50, and a 36,000-second limit. The seeded population receives 1,500 generations to find its first improvement, after which the standard 500-generation stagnation allowance applies. Every output receives the fixed-order SOCP pass described above.

\subsection{Result and Performance Analysis}
For both configurations, the percentage gap is
\[\text{Gap} = 100 \cdot \frac{\text{OurSolution} - \text{BestKnown}}{\text{BestKnown}}.\]

\subsubsection{Our algorithm as a standalone solver}

Our algorithm obtains competitive tours at much lower runtime than population-based solvers. Most gaps in Table~\ref{tab:standalone-comparison} are below $2\%$, and \texttt{bonus1000} improves the prior reported value by $1.49\%$. However, the standalone approach has limitations in handling certain instance types.

\paragraph{Standalone failure modes.}
Two structured families expose different limitations:
\begin{itemize}
  \item \textbf{RotatingDiamonds series.}
  These instances exhibit geometric regularity and low overlap, making performance highly dependent on the quality of the underlying TSP solver.
  Our implementation inherits the \(O(n \log n)\) heuristic of \textcite{Formella2024}, which yields near-linear scalability but cannot guarantee globally optimal TSP tours.
  Consequently, we observe gaps of up to $2.9\%$ (e.g., \texttt{rotatingDiamonds5}).
  Further improvement would require integrating a better TSP solver.
  \item \textbf{Bubbles series.}
  These instances feature overlapping concentric structures where the optimal CETSP tours form quasi-spiral linear traversals.
  Our solver's local search mechanisms are insufficient to escape shallow minima in such structured landscapes, and random initialization produces worse `jagged' tours.
  Performance degradation increases with instance size, reaching $8.5\%$ on \texttt{bubbles9}.
  Escaping these orderings requires search that can coordinate changes across several tour segments.
\end{itemize}

\begin{figure}[H]
  \centering
  \begin{subfigure}[b]{0.48\textwidth}
    \centering
    \includegraphics[width=\textwidth]{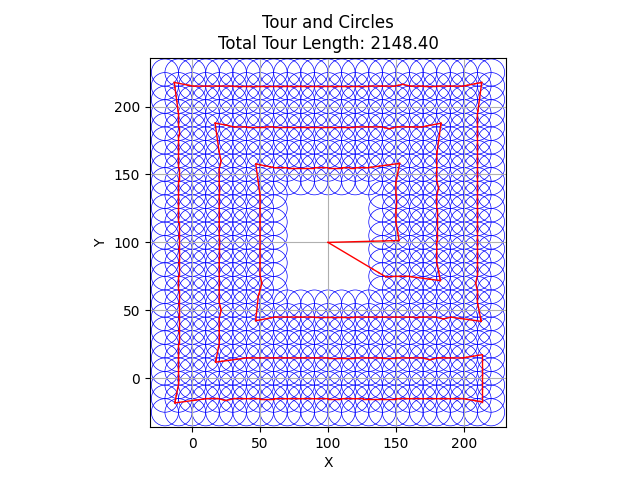}
    \caption{Benchmark solution \parencite{LeiHao2024}}
    \label{fig:bubbles9-benchmark}
  \end{subfigure}
  \hfill
  \begin{subfigure}[b]{0.48\textwidth}
    \centering
    \includegraphics[width=\textwidth]{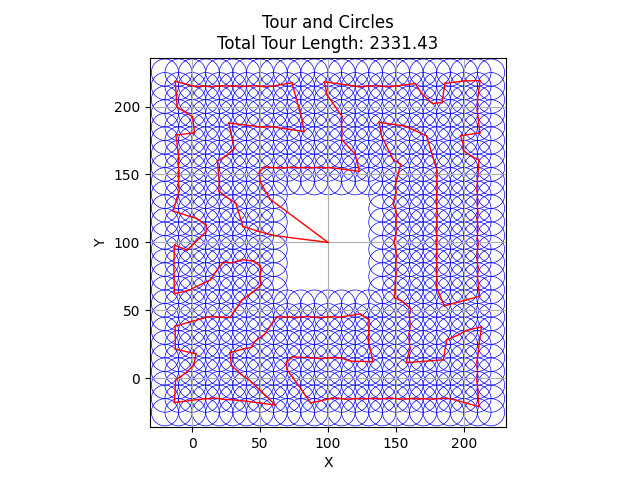}
    \caption{Our solver's solution}
    \label{fig:bubbles9-solver}
  \end{subfigure}
  \caption{
    Comparison of the benchmark and our solver's tour for the \texttt{bubbles9} instance.
    The benchmark exhibits a highly structured, spiral traversal through the concentric layers of circles, 
    whereas our solver's path shows local inefficiencies and incomplete alignment along the optimal spiral pattern.
    This case highlights the solver's difficulty escaping local minima in structured overlapping geometries.
  }
  \label{fig:bubbles9-comparison}
\end{figure}

\begin{figure}[H]
  \centering
  \begin{subfigure}[b]{0.48\textwidth}
    \centering
    \includegraphics[width=\textwidth]{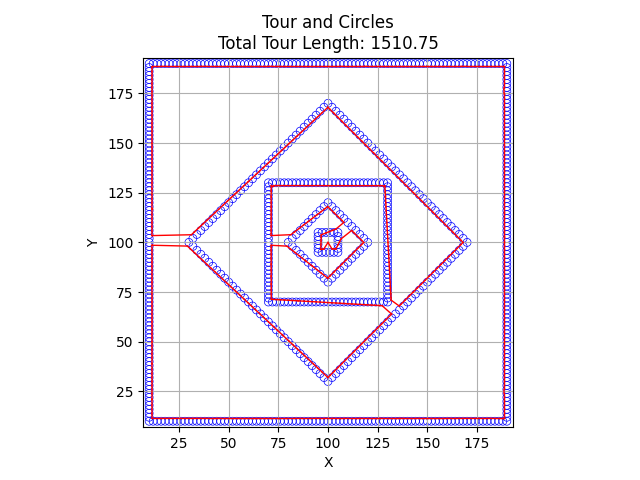}
    \caption{Benchmark solution \parencite{LeiHao2024}}
    \label{fig:rotatingDiamonds5-benchmark}
  \end{subfigure}
  \hfill
  \begin{subfigure}[b]{0.48\textwidth}
    \centering
    \includegraphics[width=\textwidth]{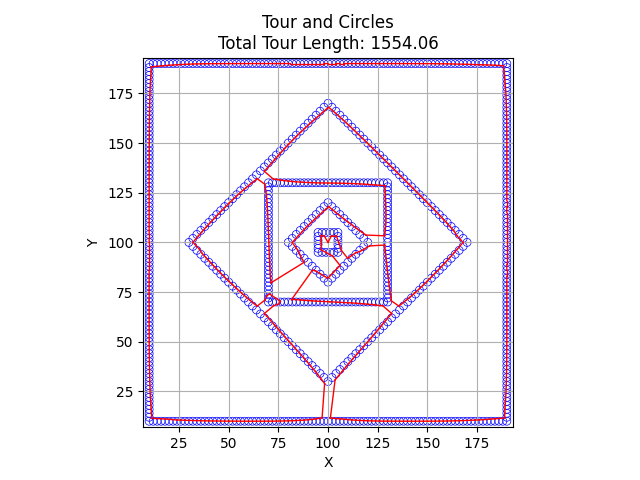}
    \caption{Our solver's solution}
    \label{fig:rotatingDiamonds5-solver}
  \end{subfigure}
  \caption{
    Comparison of the benchmark and our solver's tour for the \texttt{rotatingDiamonds5} instance.
    The benchmark achieves a better traversal order across the concentric diamond layers, 
    whereas our solver generates a worse traversal primarily due to suboptimal TSP ordering inherited from 
    the base heuristic of \textcite{Formella2024}. 
    Additional inefficiency arises from locally optimized point selections rather than globally coordinated 
    second-order cone programming (SOCP) refinements.
  }
  \label{fig:rotatingDiamonds5-comparison}
\end{figure}

\subsubsection{MA-CETSP with our initialization}

At the precision of the reference values from \textcite{LeiHao2024}, the best of 20 runs gives \textbf{17 improvements, 31 ties, and 14 losses}. Table~\ref{tab:hybrid-improvements} lists the improved values. The archived reference tours are available, but their coordinates were serialized with six significant digits. This rounding leaves 52 of the 62 exported tours with at least one visit just outside its assigned disk; the largest correction needed to restore feasibility is $6.15\times10^{-4}$ coordinate units. After repairing these serialization errors and applying fixed-order SOCP to the reference tours, all 17 hybrid tours remain shorter than their corresponding reconstructed references. Eight improvements exceed \(0.01\%\); the other nine are smaller numerical improvements.

\begin{table}[H]
\centering
\caption{Validated hybrid improvements over the best values reported by \textcite{LeiHao2024}. Improvement is positive when the hybrid tour is shorter.}
\label{tab:hybrid-improvements}
\begin{tabular}{lrrr}
\hline
Instance & Reported best & Hybrid best & Improvement \\
\hline
bonus1000 & 384.365000 & 376.667789 & 2.0026\% \\
bonus1000rdmRad & 917.610000 & 916.848440 & 0.0830\% \\
d493\_or10 & 100.721000 & 100.720299 & 0.0007\% \\
d493\_or2 & 199.015000 & 198.986227 & 0.0145\% \\
d493\_or30 & 69.757700 & 69.757626 & 0.0001\% \\
kroD100\_or10 & 89.667900 & 89.667661 & 0.0003\% \\
kroD100\_or2 & 159.037000 & 159.033834 & 0.0020\% \\
kroD100\_or30 & 58.541200 & 58.541123 & 0.0001\% \\
lin318\_or10 & 1394.630000 & 1394.617397 & 0.0009\% \\
pcb442\_or2 & 319.846000 & 319.659606 & 0.0583\% \\
pcb442\_or30 & 83.537300 & 83.537230 & 0.0001\% \\
rat195\_or10 & 67.990900 & 67.990821 & 0.0001\% \\
rat195\_or2 & 157.970000 & 157.865008 & 0.0665\% \\
rat195\_or30 & 45.701700 & 45.701642 & 0.0001\% \\
rd400\_or10 & 452.826000 & 448.455209 & 0.9652\% \\
rd400rdmRad & 1238.280000 & 1238.014237 & 0.0215\% \\
team4\_400 & 669.910000 & 668.848575 & 0.1584\% \\
\hline
\end{tabular}
\end{table}

The best-of-20 comparison measures attainable solution quality, whereas a comparison of means measures the reliability of a single independent run. Tables 5--7 of \textcite{LeiHao2024} give the mean and standard deviation of 20 runs for each instance. We compare those summary statistics with our 20 runs using two-sided Welch tests. Because the published means and standard deviations are rounded to two decimal places, each test uses the values within the corresponding rounding intervals that are least favorable to rejection. A Holm correction over all 62 tests controls the family-wise error rate at $5\%$. This yields \textbf{eight significant wins, two significant losses, and 52 inconclusive comparisons}. The full per-instance results appear in Table~\ref{tab:hybrid-comparison}.

\begin{table}[H]
\centering
\caption{Instances with a significant difference in mean performance after Holm correction. Lower tour length is better.}
\label{tab:hybrid-significance}
\begin{tabular}{lrrrl}
\hline
Instance & Hybrid mean $\pm$ SD & Lei--Hao mean $\pm$ SD & Holm $p$ & Result \\
\hline
bonus1000 & $377.337 \pm 0.374$ & $390.40 \pm 4.10$ & $7.12\!\times\!10^{-10}$ & win \\
dsj1000rdmRad & $624.604 \pm {<}0.001$ & $625.00 \pm 0.42$ & 0.0315 & win \\
lin318\_or10 & $1394.663 \pm 0.102$ & $1399.96 \pm 5.55$ & 0.0228 & win \\
lin318\_or2 & $2818.467 \pm 0.936$ & $2826.16 \pm 5.76$ & 0.000539 & win \\
pcb442\_or2 & $319.972 \pm 0.208$ & $321.05 \pm 0.97$ & 0.00549 & win \\
rat195\_or2 & $157.995 \pm 0.168$ & $158.85 \pm 0.58$ & 0.000159 & win \\
rd400\_or10 & $449.994 \pm 1.756$ & $453.96 \pm 0.99$ & $5.12\!\times\!10^{-8}$ & win \\
team5\_499 & $695.494 \pm 0.797$ & $697.37 \pm 1.68$ & 0.00668 & win \\
bubbles4 & $807.695 \pm 1.881$ & $805.37 \pm 1.70$ & 0.0121 & loss \\
bubbles7 & $1638.594 \pm 5.638$ & $1586.78 \pm 9.35$ & $2.66\!\times\!10^{-18}$ & loss \\
\hline
\end{tabular}
\end{table}

The hybrid changes the structured-instance failure profile. All five \emph{RotatingDiamonds} best results now match the reported values to practical precision. The remaining systematic failures are in the \emph{Bubbles} family. On \texttt{bubbles4}, the hybrid mean is $0.29\%$ above Lei and Hao's mean and the best is $0.22\%$ above their best; the loss is statistically significant but small. On \texttt{bubbles7}, the mean and best gaps are $3.27\%$ and $3.10\%$, respectively, giving a material and significant loss.

The larger \texttt{bubbles8} and \texttt{bubbles9} instances show a different pattern. Their best gaps are only $0.12\%$ and $0.04\%$, but their run distributions are much wider than Lei and Hao's reported distributions. Some runs escape to nearly best-known orderings while others remain in substantially worse basins. In the instrumented \texttt{bubbles9} run, the first incumbent improvement occurs at generation 891, after the ordinary 500-generation stagnation limit. This evidence is consistent with initialization-sensitive local-optimum basins rather than uniformly poor point optimization.

\begin{figure}[H]
  \centering
  \includegraphics[width=\textwidth]{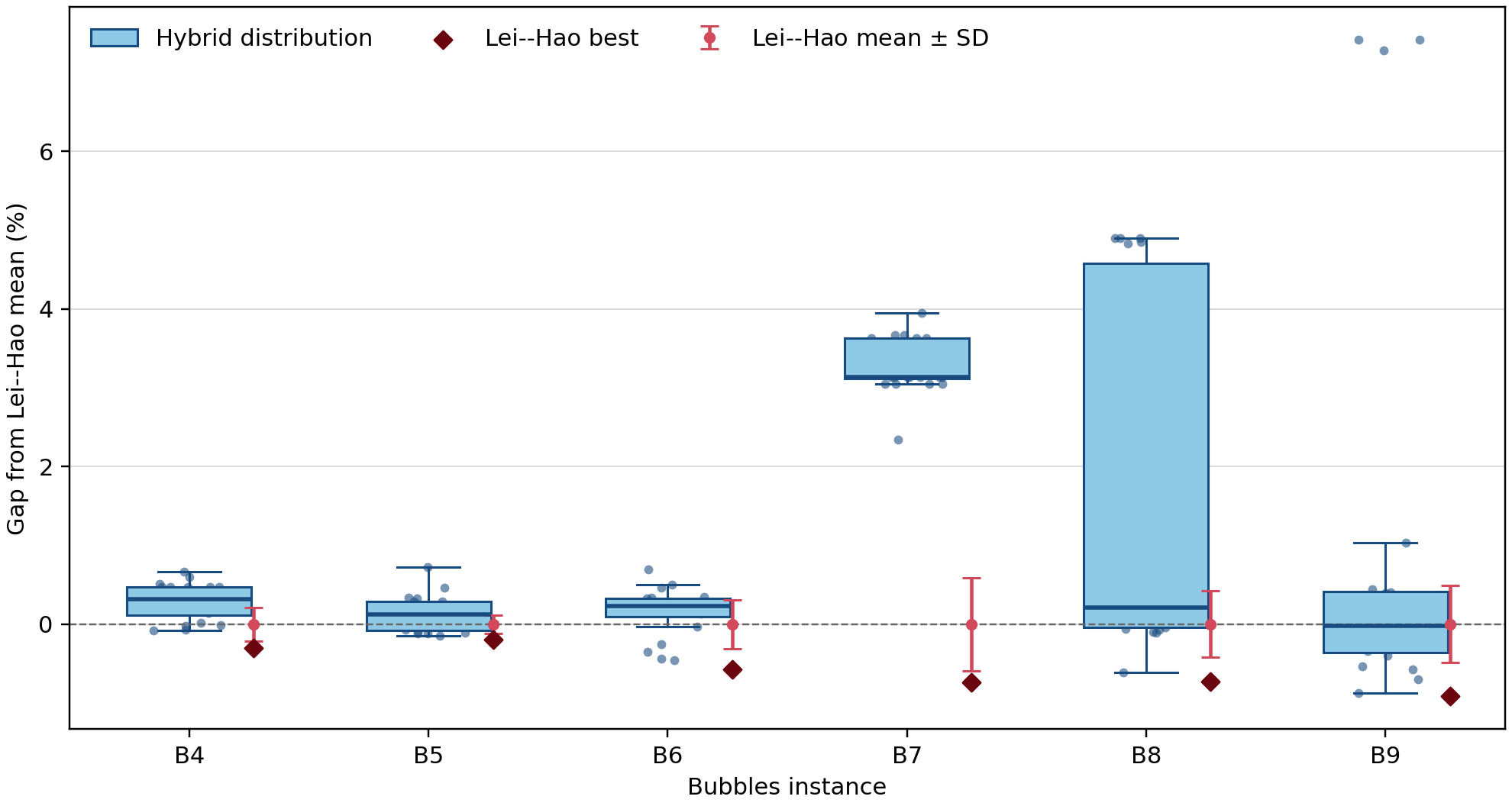}
  \caption{Run distributions for the nontrivial Bubbles instances, expressed relative to Lei and Hao's reported mean. Points show the 20 runs; boxes show their quartiles. The high variance on \texttt{bubbles8} and \texttt{bubbles9} contrasts with the persistent gap on \texttt{bubbles7}.}
  \label{fig:bubbles-hybrid-distribution}
\end{figure}

\section{Runtime Experiments}

We conducted a series of empirical runtime measurements to study the scaling behavior of our CETSP solver. 

\subsection{Random Input Experiments}

First, we generated random test instances as follows: for each of $n = 2^7, 2^8, \dots, 2^{20}$ circles, we chose the circle centers uniformly at random from the square $[-L, L]^2$, where $L$ is a problem-dependent limit, and the radii uniformly from $[0.01, 0.02] \cdot L$. Each instance was solved with 10 repetitions to average out noise.

\begin{figure}[H]
    \centering
    \includegraphics[width=0.8\textwidth]{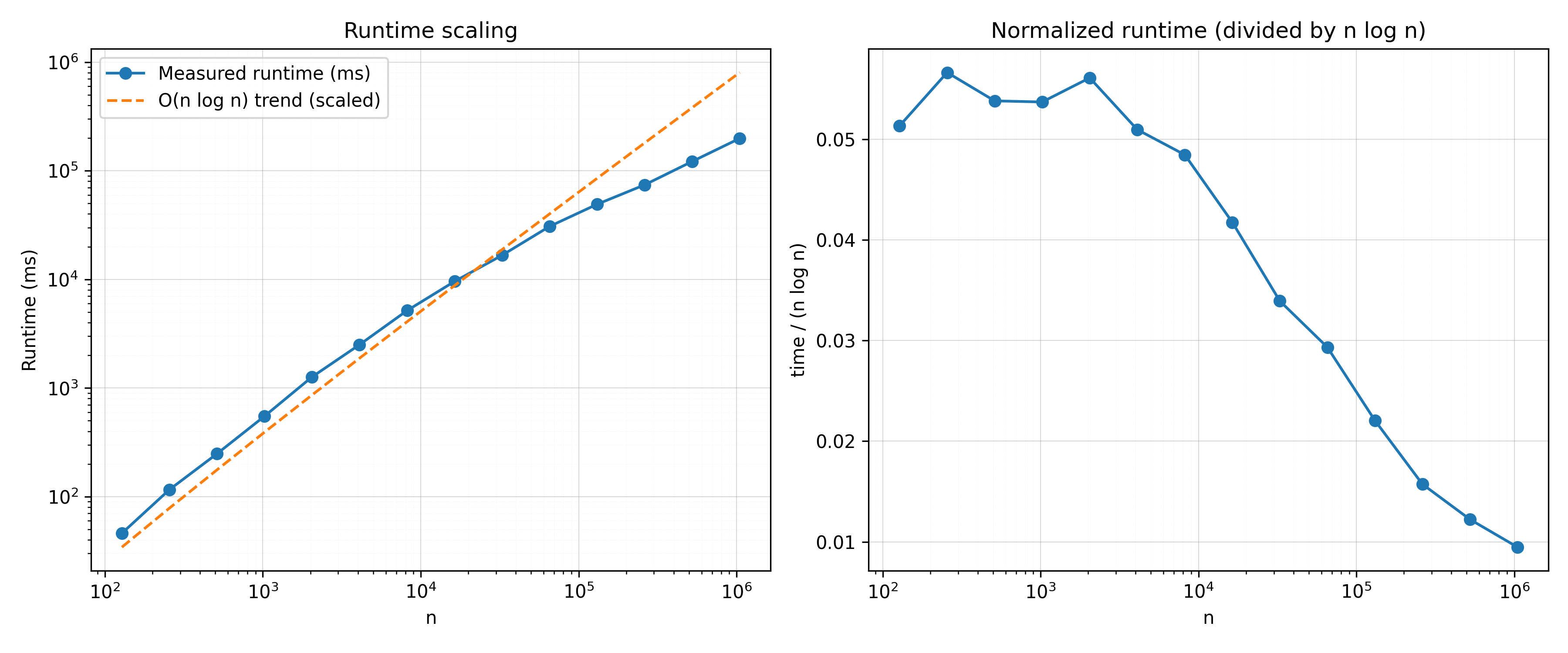}
    \caption{Runtime scaling on random CETSP instances. Runtime grows slower than $O(n \log n)$ due to sublinear growth of the tour size.}
    \label{fig:random_runtime}
\end{figure}

The results of these experiments are shown in Figure~\ref{fig:random_runtime}. Surprisingly, the measured runtime grows well below $O(n \log n)$, indicating sublinear scaling with respect to the number of circles.

Investigation of the results revealed the reason: the number of tour points increases sublinearly with the number of input circles. For example, an instance with $2^{19}$ circles produced a tour of length 20,246, while $2^{20}$ circles resulted in a tour length of 24,307. Consequently, most of the $n \log n$ component of the solver runtime is effectively scaled down by the sublinear growth of the tour.

\subsection{Structured Input Experiments}

To remove this effect and demonstrate the asymptotic scaling clearly, we generated structured test instances that guarantee a linear tour size. For each $n$, we constructed a $2$D grid of size $\lfloor \sqrt{n} \rfloor \times \lfloor \sqrt{n} \rfloor$, with centers placed at the grid points plus a small random jitter of $\pm 0.1$. Any additional points were placed uniformly at random within the grid area. The circle radii were sampled uniformly from $[0.2, 0.5]$. For these instances, we performed a single repetition of the solver for efficiency.

\begin{figure}[H]
    \centering
    \includegraphics[width=0.8\textwidth]{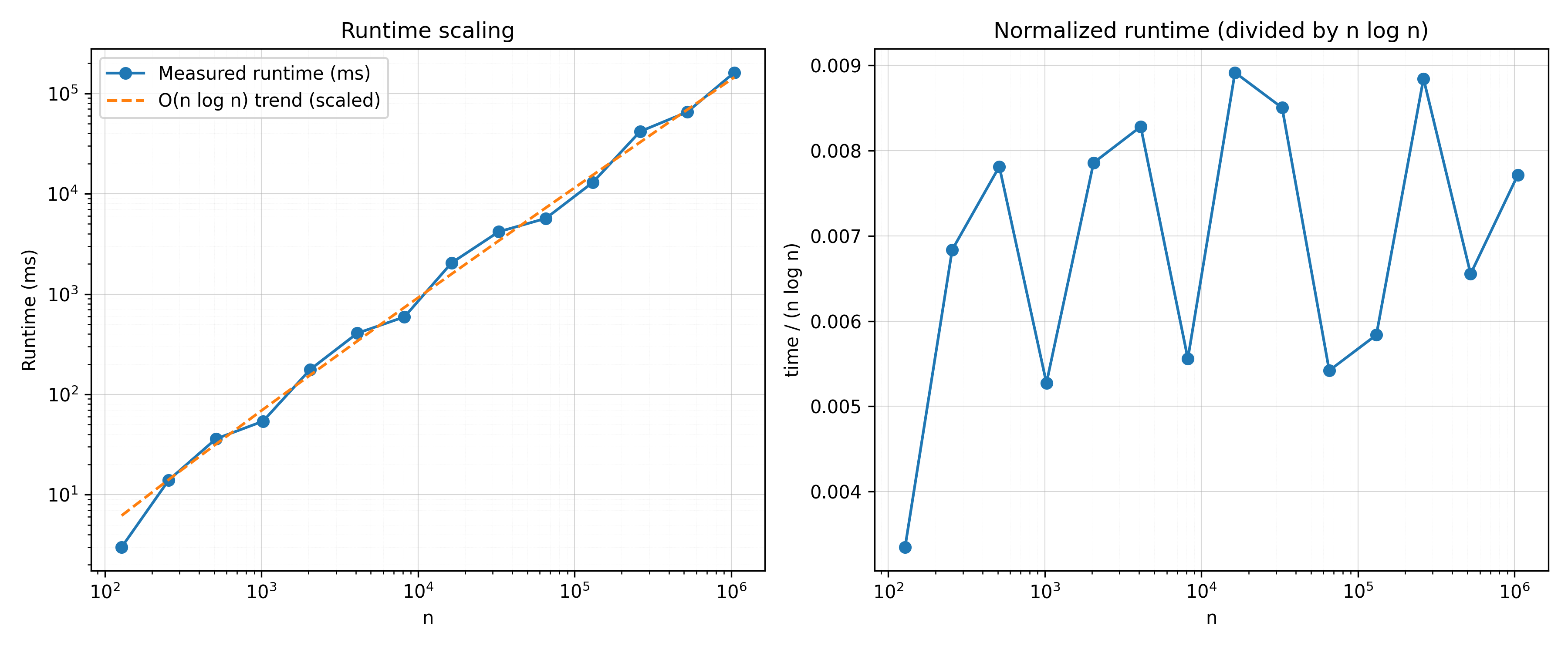}
    \caption{Runtime scaling on structured CETSP instances with linear tour size. The $O(n \log n)$ trend fits the measured data closely.}
    \label{fig:structured_runtime}
\end{figure}

The results, shown in Figure~\ref{fig:structured_runtime}, exhibit a clear $O(n \log n)$ scaling, confirming empirically that the solver runtime behaves as expected when the tour size scales linearly with $n$.

\subsection{Runtime Summary}

In summary, our experiments confirm that the solver exhibits $O(n \log n)$ runtime behavior in the regime where the tour size scales linearly with the number of input circles. On purely random instances, the observed runtime is often sub-$O(n \log n)$ due to sublinear growth of the tour size, but the structured experiments demonstrate that the underlying algorithmic complexity follows the expected scaling law.

\section{Conclusion}

The circle-based pair-center construction provides a fast standalone CETSP
heuristic and a practical source of initial populations for deeper search. The
standalone mode retains expected $O(n\log n)$ behavior and exposes failures
caused by globally poor visit orders. The hybrid mode substantially closes
those gaps: its best-of-20 results improve 17 values reported by
\textcite{LeiHao2024}, and its mean performance gives eight significant wins
and two losses after correction across the benchmark suite.

The hybrid does not remove every ordering failure. In particular,
\texttt{bubbles7} remains trapped in a consistently inferior basin, while the
wide distributions on \texttt{bubbles8} and \texttt{bubbles9} show that good
basins can be reached but not reliably. A matched ablation of native MA-CETSP
initialization, construction seeding, fan-out, and the extended initial
stagnation allowance is the next step for isolating which parts of the pipeline
produce the observed gains.

\pagebreak
\printbibliography
\pagebreak

\appendix
\section{Appendix: Table of Results}

\begin{longtable}{lrrrrr}
\caption{Standalone construction results for all instances} \label{tab:standalone-comparison} \\
\hline
Instance & N & OR\% & Benchmark Value & Output Value\footnote{Output values obtained using our CETSP solver over 1000 runs, the best value is reported.} & Gap\% \\
\hline
\endfirsthead

\multicolumn{6}{c}{{\bfseries \tablename\ \thetable{} -- continued from previous page}} \\
\hline
Instance & N & OR\% & Benchmark Value & Output Value & Gap\% \\
\hline
\endhead

\hline \multicolumn{6}{r}{{Continued on next page}} \\
\endfoot

\hline
\endlastfoot
bonus1000\footnote{The \texttt{bonus1000} test case was ran 10000 times to improve the best value.} & 1000 & 12.26 & 384.365 & 378.622 & -1.49 \\
bonus1000rdmRad & 1000 & 6.13 & 917.610 & 937.247 & 2.14 \\
bubbles1 & 36 & 11.11 & 349.135 & 349.255 & 0.03 \\
bubbles2 & 76 & 9.09 & 428.279 & 428.368 & 0.02 \\
bubbles3 & 126 & 7.69 & 529.955 & 531.243 & 0.24 \\
bubbles4 & 184 & 6.67 & 802.974 & 809.216 & 0.78 \\
bubbles5 & 250 & 5.88 & 1035.320 & 1055.488 & 1.95 \\
bubbles6 & 324 & 5.26 & 1220.070 & 1301.459 & 6.67 \\
bubbles7 & 406 & 4.76 & 1575.040 & 1647.837 & 4.62 \\
bubbles8 & 496 & 4.35 & 1881.930 & 1995.234 & 6.02 \\
bubbles9 & 594 & 4.00 & 2148.400 & 2331.431 & 8.52 \\
chaoSingleDep & 200 & 2.84 & 1039.610 & 1042.864 & 0.31 \\
concentricCircles1 & 16 & 15.00 & 53.158 & 53.201 & 0.08 \\
concentricCircles2 & 36 & 7.50 & 153.132 & 153.870 & 0.48 \\
concentricCircles3 & 60 & 5.00 & 270.007 & 271.744 & 0.64 \\
concentricCircles4 & 104 & 3.75 & 451.870 & 460.079 & 1.82 \\
concentricCircles5 & 148 & 3.00 & 632.977 & 639.080 & 0.96 \\
d493\_or10 & 493 & 10.00 & 100.721 & 100.959 & 0.24 \\
d493\_or2 & 493 & 2.00 & 199.015 & 200.798 & 0.90 \\
d493\_or30 & 493 & 30.00 & 69.758 & 69.758 & 0.00 \\
d493rdmRad & 493 & 13.44 & 134.226 & 134.621 & 0.29 \\
dsj1000\_or10 & 1000 & 10.00 & 373.759 & 376.445 & 0.72 \\
dsj1000\_or2 & 1000 & 2.00 & 909.502 & 922.440 & 1.42 \\
dsj1000\_or30 & 1000 & 30.00 & 199.948 & 200.037 & 0.04 \\
dsj1000rdmRad & 1000 & 12.47 & 624.604 & 626.148 & 0.25 \\
kroD100\_or10 & 100 & 10.00 & 89.668 & 89.781 & 0.13 \\
kroD100\_or2 & 100 & 2.00 & 159.037 & 159.603 & 0.36 \\
kroD100\_or30 & 100 & 30.00 & 58.541 & 58.576 & 0.06 \\
kroD100rdmRad & 100 & 4.03 & 141.829 & 142.199 & 0.26 \\
lin318\_or10 & 318 & 10.00 & 1394.630 & 1404.343 & 0.70 \\
lin318\_or2 & 318 & 2.00 & 2816.580 & 2833.410 & 0.60 \\
lin318\_or30 & 318 & 30.00 & 765.964 & 765.998 & 0.00 \\
lin318rdmRad & 318 & 10.05 & 2047.110 & 2052.486 & 0.26 \\
pcb442\_or10 & 442 & 10.00 & 142.573 & 143.302 & 0.51 \\
pcb442\_or2 & 442 & 2.00 & 319.846 & 332.701 & 4.02 \\
pcb442\_or30 & 442 & 30.00 & 83.537 & 83.538 & 0.00 \\
pcb442rdmRad & 442 & 9.39 & 219.219 & 220.280 & 0.48 \\
rat195\_or10 & 195 & 10.00 & 67.991 & 68.080 & 0.13 \\
rat195\_or2 & 195 & 2.00 & 157.970 & 160.495 & 1.60 \\
rat195\_or30 & 195 & 30.00 & 45.702 & 45.702 & 0.00 \\
rat195rdmRad & 195 & 42.60 & 68.224 & 68.252 & 0.04 \\
rd400\_or10 & 400 & 10.00 & 452.826 & 453.943 & 0.25 \\
rd400\_or2 & 400 & 2.00 & 1018.460 & 1040.517 & 2.17 \\
rd400\_or30 & 400 & 30.00 & 224.839 & 224.910 & 0.03 \\
rd400rdmRad & 400 & 0.99 & 1238.280 & 1279.515 & 3.33 \\
rotatingDiamonds1 & 20 & 20.00 & 32.389 & 32.431 & 0.13 \\
rotatingDiamonds2 & 60 & 5.00 & 140.477 & 140.636 & 0.11 \\
rotatingDiamonds3 & 180 & 3.33 & 380.882 & 381.475 & 0.16 \\
rotatingDiamonds4 & 320 & 1.43 & 770.661 & 790.601 & 2.59 \\
rotatingDiamonds5 & 680 & 1.11 & 1510.750 & 1554.064 & 2.87 \\
team1\_100 & 100 & 9.33 & 307.337 & 308.339 & 0.33 \\
team1\_100rdmRad & 100 & 7.69 & 388.537 & 399.272 & 2.76 \\
team2\_200 & 200 & 20.06 & 246.683 & 247.401 & 0.29 \\
team2\_200rdmRad & 200 & 5.19 & 613.659 & 618.001 & 0.71 \\
team3\_300 & 300 & 7.02 & 461.889 & 462.960 & 0.23 \\
team3\_300rdmRad & 300 & 23.70 & 378.087 & 379.137 & 0.28 \\
team4\_400 & 400 & 5.01 & 669.910 & 674.344 & 0.66 \\
team4\_400rdmRad & 400 & 2.05 & 984.240 & 1013.142 & 2.94 \\
team5\_499 & 499 & 2.00 & 693.798 & 704.900 & 1.60 \\
team5\_499rdmRad & 499 & 20.14 & 446.191 & 446.525 & 0.07 \\
team6\_500 & 500 & 27.06 & 225.216 & 225.451 & 0.10 \\
team6\_500rdmRad & 500 & 10.02 & 620.886 & 623.678 & 0.45 \\
\end{longtable}

\begingroup
\footnotesize
\begin{longtable}{lrrrrl}
\caption{Hybrid results for all instances. A negative gap denotes an
improvement over the reported best. The final column gives the direction only
for mean differences significant after Holm correction; ``--'' denotes an
inconclusive comparison.} \label{tab:hybrid-comparison} \\
\hline
Instance & Reported best & Hybrid best & Gap\% & Hybrid mean $\pm$ SD & Holm \\
\hline
\endfirsthead

\multicolumn{6}{c}{{\bfseries \tablename\ \thetable{} -- continued from previous page}} \\
\hline
Instance & Reported best & Hybrid best & Gap\% & Hybrid mean $\pm$ SD & Holm \\
\hline
\endhead

\hline \multicolumn{6}{r}{{Continued on next page}} \\
\endfoot

\hline
\endlastfoot
bonus1000 & 384.365 & 376.667789 & -2.0026 & $377.337 \pm 0.374$ & win \\
bonus1000rdmRad & 917.61 & 916.848440 & -0.0830 & $919.024 \pm 1.707$ & -- \\
bubbles1 & 349.135 & 349.134889 & 0.0000 & $349.135 \pm 0.000$ & -- \\
bubbles2 & 428.279 & 428.279256 & 0.0001 & $428.279 \pm 0.000$ & -- \\
bubbles3 & 529.955 & 529.954758 & 0.0000 & $529.992 \pm 0.083$ & -- \\
bubbles4 & 802.974 & 804.720672 & 0.2175 & $807.695 \pm 1.881$ & loss \\
bubbles5 & 1035.32 & 1035.884301 & 0.0545 & $1038.761 \pm 2.364$ & -- \\
bubbles6 & 1220.07 & 1221.565344 & 0.1226 & $1229.065 \pm 3.824$ & -- \\
bubbles7 & 1575.04 & 1623.889080 & 3.1015 & $1638.594 \pm 5.638$ & loss \\
bubbles8 & 1881.93 & 1884.133589 & 0.1171 & $1928.292 \pm 43.592$ & -- \\
bubbles9 & 2148.4 & 2149.341438 & 0.0438 & $2190.645 \pm 59.965$ & -- \\
chaoSingleDep & 1039.61 & 1039.610891 & 0.0001 & $1039.611 \pm 0.001$ & -- \\
concentricCircles1 & 53.158 & 53.157996 & 0.0000 & $53.158 \pm 0.000$ & -- \\
concentricCircles2 & 153.132 & 153.132283 & 0.0002 & $153.159 \pm 0.016$ & -- \\
concentricCircles3 & 270.007 & 270.013338 & 0.0023 & $270.192 \pm 0.178$ & -- \\
concentricCircles4 & 451.87 & 452.081784 & 0.0469 & $452.986 \pm 0.831$ & -- \\
concentricCircles5 & 632.977 & 633.050759 & 0.0117 & $633.690 \pm 0.514$ & -- \\
d493\_or10 & 100.721 & 100.720299 & -0.0007 & $100.720 \pm 0.000$ & -- \\
d493\_or2 & 199.015 & 198.986227 & -0.0145 & $199.590 \pm 0.442$ & -- \\
d493\_or30 & 69.7577 & 69.757626 & -0.0001 & $69.758 \pm 0.000$ & -- \\
d493rdmRad & 134.226 & 134.226358 & 0.0003 & $134.226 \pm 0.000$ & -- \\
dsj1000\_or10 & 373.759 & 373.758904 & 0.0000 & $373.787 \pm 0.048$ & -- \\
dsj1000\_or2 & 909.502 & 909.795943 & 0.0323 & $915.517 \pm 5.337$ & -- \\
dsj1000\_or30 & 199.948 & 199.947960 & 0.0000 & $199.948 \pm 0.000$ & -- \\
dsj1000rdmRad & 624.604 & 624.604269 & 0.0000 & $624.604 \pm 0.000$ & win \\
kroD100\_or10 & 89.6679 & 89.667661 & -0.0003 & $89.668 \pm 0.000$ & -- \\
kroD100\_or2 & 159.037 & 159.033834 & -0.0020 & $159.034 \pm 0.000$ & -- \\
kroD100\_or30 & 58.5412 & 58.541123 & -0.0001 & $58.541 \pm 0.000$ & -- \\
kroD100rdmRad & 141.829 & 141.828873 & -0.0001 & $141.829 \pm 0.000$ & -- \\
lin318\_or10 & 1394.63 & 1394.617397 & -0.0009 & $1394.663 \pm 0.102$ & win \\
lin318\_or2 & 2816.58 & 2817.020280 & 0.0156 & $2818.467 \pm 0.936$ & win \\
lin318\_or30 & 765.964 & 765.963925 & 0.0000 & $765.964 \pm 0.000$ & -- \\
lin318rdmRad & 2047.11 & 2047.106469 & -0.0002 & $2047.106 \pm 0.000$ & -- \\
pcb442\_or10 & 142.573 & 142.592210 & 0.0135 & $142.685 \pm 0.048$ & -- \\
pcb442\_or2 & 319.846 & 319.659606 & -0.0583 & $319.972 \pm 0.208$ & win \\
pcb442\_or30 & 83.5373 & 83.537230 & -0.0001 & $83.537 \pm 0.000$ & -- \\
pcb442rdmRad & 219.219 & 219.218830 & -0.0001 & $219.301 \pm 0.171$ & -- \\
rat195\_or10 & 67.9909 & 67.990821 & -0.0001 & $67.991 \pm 0.001$ & -- \\
rat195\_or2 & 157.97 & 157.865008 & -0.0665 & $157.995 \pm 0.168$ & win \\
rat195\_or30 & 45.7017 & 45.701642 & -0.0001 & $45.702 \pm 0.000$ & -- \\
rat195rdmRad & 68.224 & 68.224003 & 0.0000 & $68.224 \pm 0.000$ & -- \\
rd400\_or10 & 452.826 & 448.455209 & -0.9652 & $449.994 \pm 1.756$ & win \\
rd400\_or2 & 1018.46 & 1018.464316 & 0.0004 & $1020.643 \pm 1.286$ & -- \\
rd400\_or30 & 224.839 & 224.838597 & -0.0002 & $224.839 \pm 0.000$ & -- \\
rd400rdmRad & 1238.28 & 1238.014237 & -0.0215 & $1239.592 \pm 2.033$ & -- \\
rotatingDiamonds1 & 32.389 & 32.389030 & 0.0001 & $32.389 \pm 0.000$ & -- \\
rotatingDiamonds2 & 140.477 & 140.477409 & 0.0003 & $140.477 \pm 0.000$ & -- \\
rotatingDiamonds3 & 380.882 & 380.882434 & 0.0001 & $380.883 \pm 0.000$ & -- \\
rotatingDiamonds4 & 770.661 & 770.660642 & 0.0000 & $770.690 \pm 0.065$ & -- \\
rotatingDiamonds5 & 1510.75 & 1510.748945 & -0.0001 & $1510.790 \pm 0.064$ & -- \\
team1\_100 & 307.337 & 307.336869 & 0.0000 & $307.337 \pm 0.000$ & -- \\
team1\_100rdmRad & 388.537 & 388.536876 & 0.0000 & $388.537 \pm 0.000$ & -- \\
team2\_200 & 246.683 & 246.682984 & 0.0000 & $246.683 \pm 0.000$ & -- \\
team2\_200rdmRad & 613.659 & 613.659481 & 0.0001 & $614.473 \pm 1.550$ & -- \\
team3\_300 & 461.889 & 461.888593 & -0.0001 & $461.905 \pm 0.028$ & -- \\
team3\_300rdmRad & 378.087 & 378.087142 & 0.0000 & $378.087 \pm 0.000$ & -- \\
team4\_400 & 669.91 & 668.848575 & -0.1584 & $673.820 \pm 2.246$ & -- \\
team4\_400rdmRad & 984.24 & 984.320986 & 0.0082 & $985.145 \pm 0.755$ & -- \\
team5\_499 & 693.798 & 694.272958 & 0.0685 & $695.494 \pm 0.797$ & win \\
team5\_499rdmRad & 446.191 & 446.191187 & 0.0000 & $446.191 \pm 0.000$ & -- \\
team6\_500 & 225.216 & 225.216005 & 0.0000 & $225.216 \pm 0.000$ & -- \\
team6\_500rdmRad & 620.886 & 620.885859 & 0.0000 & $621.004 \pm 0.292$ & -- \\
\end{longtable}
\endgroup

\begingroup
\small
\begin{longtable}{lrrrr}
\caption{Standalone runtime comparison. Standalone times were measured on an
AMD Ryzen 9 7940HS and scaled to the AMD Opteron 4184 used by
\textcite{LeiHao2024} with a PassMark-based factor of 10.424. Ratio is adjusted
standalone time divided by reference time.} \label{tab:runtime_adjusted} \\
\hline
Instance & Reference (s) & Standalone (s) & Adjusted (s) & Ratio \\
\hline
\endfirsthead

\multicolumn{5}{c}{{\bfseries \tablename\ \thetable{} -- continued from previous page}} \\
\hline
Instance & Reference (s) & Standalone (s) & Adjusted (s) & Ratio \\
\hline
\endhead

\hline \multicolumn{5}{r}{{Continued on next page}} \\
\endfoot

\hline
\endlastfoot

bonus1000 & 2016.81 & 0.300 & 3.127 & 0.0016 \\
bonus1000rdmRad & 890.66 & 0.239 & 2.491 & 0.0028 \\
bubbles1 & 31.12 & 0.014 & 0.146 & 0.0047 \\
bubbles2 & 41.73 & 0.025 & 0.261 & 0.0062 \\
bubbles3 & 193.23 & 0.051 & 0.532 & 0.0028 \\
bubbles4 & 173.63 & 0.077 & 0.803 & 0.0046 \\
bubbles5 & 247.52 & 0.125 & 1.303 & 0.0053 \\
bubbles6 & 358.74 & 0.192 & 2.001 & 0.0056 \\
bubbles7 & 650.29 & 0.228 & 2.377 & 0.0037 \\
bubbles8 & 790.10 & 0.292 & 3.044 & 0.0039 \\
bubbles9 & 893.73 & 0.443 & 4.618 & 0.0052 \\
chaoSingleDep & 76.44 & 0.059 & 0.615 & 0.0080 \\
concentricCircles1 & 29.31 & 0.002 & 0.021 & 0.0007 \\
concentricCircles2 & 139.32 & 0.011 & 0.115 & 0.0008 \\
concentricCircles3 & 67.01 & 0.023 & 0.240 & 0.0036 \\
concentricCircles4 & 193.29 & 0.049 & 0.511 & 0.0026 \\
concentricCircles5 & 208.92 & 0.082 & 0.855 & 0.0041 \\
d493\_or10 & 352.52 & 0.033 & 0.344 & 0.0010 \\
d493\_or2 & 689.44 & 0.140 & 1.459 & 0.0021 \\
d493\_or30 & 116.74 & 0.267 & 2.783 & 0.0238 \\
d493rdmRad & 71.85 & 0.189 & 1.970 & 0.0274 \\
dsj1000\_or10 & 942.49 & 0.089 & 0.928 & 0.0010 \\
dsj1000\_or2 & 2578.76 & 0.240 & 2.502 & 0.0010 \\
dsj1000\_or30 & 249.61 & 0.597 & 6.223 & 0.0249 \\
dsj1000rdmRad & 176.72 & 0.239 & 2.491 & 0.0141 \\
kroD100\_or10 & 50.76 & 0.030 & 0.313 & 0.0062 \\
kroD100\_or2 & 158.32 & 0.020 & 0.208 & 0.0013 \\
kroD100\_or30 & 39.25 & 0.038 & 0.396 & 0.0101 \\
kroD100rdmRad & 158.19 & 0.009 & 0.094 & 0.0006 \\
lin318\_or10 & 193.10 & 0.047 & 0.490 & 0.0025 \\
lin318\_or2 & 327.54 & 0.094 & 0.981 & 0.0030 \\
lin318\_or30 & 71.47 & 0.199 & 2.076 & 0.0290 \\
lin318rdmRad & 66.44 & 0.059 & 0.615 & 0.0093 \\
pcb442\_or10 & 467.38 & 0.082 & 0.855 & 0.0018 \\
pcb442\_or2 & 1334.95 & 0.033 & 0.344 & 0.0003 \\
pcb442\_or30 & 117.86 & 0.140 & 1.459 & 0.0124 \\
pcb442rdmRad & 160.56 & 0.267 & 2.783 & 0.0173 \\
rat195\_or10 & 170.90 & 0.189 & 1.970 & 0.0115 \\
rat195\_or2 & 322.37 & 0.089 & 0.928 & 0.0029 \\
rat195\_or30 & 50.19 & 0.240 & 2.502 & 0.0499 \\
rat195rdmRad & 32.17 & 0.597 & 6.223 & 0.1934 \\
rd400\_or10 & 540.53 & 0.239 & 2.491 & 0.0046 \\
rd400\_or2 & 450.63 & 0.030 & 0.313 & 0.0007 \\
rd400\_or30 & 91.67 & 0.020 & 0.208 & 0.0023 \\
rd400rdmRad & 635.69 & 0.038 & 0.396 & 0.0006 \\
rotatingDiamonds1 & 33.70 & 0.059 & 0.615 & 0.0182 \\
rotatingDiamonds2 & 46.81 & 0.082 & 0.855 & 0.0183 \\
rotatingDiamonds3 & 87.69 & 0.105 & 1.094 & 0.0125 \\
rotatingDiamonds4 & 229.92 & 0.302 & 3.149 & 0.0137 \\
rotatingDiamonds5 & 597.20 & 0.073 & 0.761 & 0.0013 \\
team1\_100 & 54.29 & 0.009 & 0.094 & 0.0017 \\
team1\_100rdmRad & 38.02 & 0.050 & 0.521 & 0.0137 \\
team2\_200 & 181.62 & 0.122 & 1.270 & 0.0070 \\
team2\_200rdmRad & 93.86 & 0.022 & 0.230 & 0.0024 \\
team3\_300 & 435.69 & 0.287 & 2.992 & 0.0069 \\
team3\_300rdmRad & 48.46 & 0.133 & 1.389 & 0.0287 \\
team4\_400 & 648.08 & 0.273 & 2.844 & 0.0044 \\
team4\_400rdmRad & 734.98 & 0.055 & 0.574 & 0.0008 \\
team5\_499 & 561.82 & 0.003 & 0.031 & 0.0001 \\
team5\_499rdmRad & 51.36 & 0.026 & 0.271 & 0.0053 \\
team6\_500 & 208.22 & 0.114 & 1.188 & 0.0057 \\
team6\_500rdmRad & 261.21 & 0.229 & 2.384 & 0.0091 \\
\end{longtable}
\endgroup

\end{document}